\documentclass[fleqn,usenatbib,useAMS]{mnras}

\usepackage{newtxtext,newtxmath}

\usepackage{graphicx}	
\usepackage{amsmath}	
\usepackage{multicol}
\usepackage{longtable}
\usepackage{bm}	
\usepackage{pdflscape}

\let\VANthebibliography\thebibliography
\def\thebibliography{\DeclareRobustCommand{\VAN}[3]{##3}\VANthebibliography}
\usepackage[dvipsnames]{xcolor}
\usepackage[T1]{fontenc}
\usepackage{ae,aecompl}
\usepackage{newtxtext,newtxmath}
\usepackage[normalem]{ulem}
\usepackage{soul}
\usepackage{cancel}

\title[$AstroSat$ observation of GX 340+0.]{ Spectral and Timing evolution of GX 340+0 along its Z-track}

\author[Chattopadhyay et al.]{
Suchismito Chattopadhyay$^{1}$\thanks{suchismitochattopadhyay@gmail.com},
Yashpal Bhulla$^{2}$\thanks{yash.pkn@gmail.com},
Ranjeev Misra$^{3}$ \thanks{rmisra@iucaa.in},
and Soma Mandal$^{1}$\thanks{soma2778.wbes@gmail.com}
\\
$^{1}$Department of Physics, Government Girls' General Degree College,7 Mayurbhanj Road, Kolkata-700023, India\\
$^{2}$Pacific Academy of Higher Education and Research University, Udaipur-313003, India\\
$^{3}$IUCAA, Post Bag 4, Ganeshkhind , Pune , Maharashtra-411007, India\\
}

\date{Accepted 2024 February 02. Received 2024 February 02; in original form 2023 February 21}

\pubyear{2022}

\begin{document}
\label{firstpage}
\pagerange{\pageref{firstpage}--\pageref{lastpage}}
\maketitle

\begin{abstract}
We present the results from spectral and timing study of the Z source GX 340+0 using {\it AstroSat}'s SXT and LAXPC data. During the observation the source traced out the complete Z-track, allowing for the spectral evolution study of the Horizontal, Normal and Flaring branches (HB, NB and FB) as well as the hard and soft apexes (HA and SA). The spectra are better and more physically described by a blackbody component and a hot Comptonizing corona with a varying covering fraction, rather than one having a disc component. Along the track, the Comptonized flux (as well as the covering fraction) monotonically decreases. It is the blackbody component (both the temperature and radius) which varies non-monotonically and hence gives rise to the Z-track behaviour. Rapid timing study reveals a prominent Quasi-periodic Oscillation (QPO) at $\sim 50$ Hz at the HB, HA and upper NB, while a QPO at $\sim 6$ Hz is seen for the other branches. The fractional r.m.s of the QPOs increase with energy and exhibit soft lags in all branches except SA and FB.
	
\end{abstract}




\section{Introduction}

In accreting neutron star low mass X-ray binaries (NS-LMXB), the matter from low mass companion stars [$\leq M_{\odot}$] gets accreted through the inner Lagrangian point via Roche-lobe overflow mechanism 
\citep{klitzing:frank1985accretion}. In several NS-LMXBs sources, the radiating luminosity  exceeds $10^{36}-10^{38}$ ergs/sec i.e. their luminosity is close to the Eddington limit $L_{edd}$\footnote{For NSLMXBs value of $L_{edd}$ is approximately $1.26\times 10^{38}$ ergs/sec \citep{klitzing:frank1985accretion}}
\citep{klitzing:1990ARA&A..28..183C,klitzing:2006csxs.book..623T,klitzing:Wang2016ABR}.\\
Weakly magnetized  neutron stars ($10^8-10^{10} G$) are classified into two categories named as Atoll (L$\approxeq 0.001-0.1 L_{edd}$) and Z sources (L$\approxeq 0.5-1 L_{edd}$) \citep{klitzing:hasinger1989two,klitzing:2021ASSL..461..263M}. The classification is based on the tracks made in the hardness-intensity-diagram (HID) or the colour-colour diagram (CCD).  Z-track sources usually cover the track on a short time scale (a few hours to weeks) as compared to atoll sources (a few weeks to months) \citep{klitzing:2002ApJ...567.1091P,klitzing:Lin_2009}. 
The Z-track sources are expanded on the three different branches in HID: Horizontal branch(HB), Normal branch (NB), Flaring branch(FB) \citep{klitzing:hasinger1989two,klitzing:1989ASIC..262..295S}. The mass accretion rate maybe increasing from HB to FB \citep{klitzing:priedhorsky1986bimodal}, but disagreement exists regarding this point of view \citep{klitzing:2009MNRAS.398.1352A,klitzing:2010A&A...512A...9B}. Z sources are further categorized into two sub-groups referred to as Cyg-like sources and Sco-like sources. Cyg-like sources show the presence of a long HB branch, whereas, in Sco-like sources, a more prominent FB than the HB has been observed \citep{klitzing:2006A&A...460..233C}.\\
The X-ray emission spectra of Z-track sources have been described by two different scenarios or continuum models. First, a multicolour blackbody emission component from the disc is included with a thermal comptonized component from a hot corona region or boundary layer \citep{klitzing:mitsuda1988qpo, klitzing:hanawa1989x, klitzing:2002MNRAS.331..453D, klitzing:2003A&A...410..217G, klitzing:2003MNRAS.346..933A}. In the second scenario, a blackbody emission from the surface of the  NS or boundary layer is used with a comptonized spectral emission from a hot extended corona \citep{klitzing:Smale2001,klitzing:2003A&A...405..237B,klitzing:2006A&A...460..233C, klitzing:2010A&A...512A...9B, klitzing:Church:2012nb}. In both scenarios, spectral parameters are seen to change along the Z-track of the source. In another interpretation, the emission from a multi-colour disc and a blackbody component are simultaneously dominant in the spectra of NS-LMXBs \citep{klitzing:2018ApJ...867...64C}.\\
The rapid temporal behaviour of these sources depends on the position they have in the Z-track. Quasi-periodic Oscillations (QPO) have been observed in their power spectra. The QPO frequency in the horizontal branch or hard apex is seen in the range 20-60 Hz \citep{klitzing:1990A&A...235..162V, klitzing:Kuulkers1997} named as HBO, whereas normal branch oscillation ranges over 5-20 Hz \citep{klitzing:1989ASIC..262..295S,klitzing:1992MNRAS.256..545L, klitzing:1994A&A...289..795K, klitzing:Kuulkers1997}. Apart from these QPOs, twin kHz-QPOs are typically observed in  Z-sources discovered  \citep{klitzing:1997ApJ...490L.157W,klitzing:1998ApJ...504L..35W,klitzing:1998ApJ...500L.167Z,klitzing:1998ApJ...493L..87W,klitzing:2000MNRAS.318..938M}.\\
GX 340+0 is a Cyg-like source which was discovered in 1967 as a Galactic X-ray source by  Friedman, Byram \& Chubb and using {\it Aerobee} rocket data \citet{klitzing:margon1971evidence} determined it as a neutron star binary.\citet{klitzing:10.1046/j.1365-8711.2000.03443.x} estimated the distance of the source to be  11$\pm$3 kpc. The source shows a clear distinctive presence of three main branches while the existence of an extra fourth branch trailing the FB of the source has been reported\citep{klitzing:1998ApJ...499L.191J}. The energy spectrum and the power density spectrum have been previously studied using   $EXOSAT$ data \citep{klitzing:1996A&A...314..567K} and {\it RXTE} \citep{klitzing:Jonker_2000}.
The HBO frequency of the source ranges over 32Hz - 50Hz \citep{klitzing:10.1093/mnras/249.1.113} while the NBO one varies between 5-7 Hz. Twin kHz-QPOs were also found in the HB, with the centroid frequency (lower kHz QPO)  247$\pm$6 Hz and (upper kHz QPO) 567 $\pm$ 39 Hz \citep{klitzing:1998ApJ...499L.191J}. With $EXOSAT$ data \citet{klitzing:schulz1993compton} showed the spectra can be modelled by a simple comptonizing model in 2-12 keV energy band. Using $BeppoSAX$ data (0.1-30 keV) \citet{klitzing:Iaria_2006} fitted the spectra of the source by the blackbody and comptonization component. \citet{klitzing:2006A&A...460..233C} analyzed the $RXTE$ data of GX 340+0 and modelled the spectra with blackbody emission from the boundary-layer with an extended corona producing the thermal Comptonization component. Thus, spectral analysis of the source preferred the interpretation that there is a blackbody component instead of a disc one.\\
The SXT \citep{klitzing:singh2014astrosat, klitzing:yadav2016large, klitzing:2017JApA...38...29S} and LAXPC \citep{klitzing:Yadav_2017, klitzing:2017CSci..113..591Y, klitzing:Agrawal_2017, klitzing:Antia_2021} instruments onboard {\it AstroSat} \citep{klitzing:2014SPIE.9144E..1SS} provide broad band spectral data and at the same time LAXPC provides rapid temporal information. This provides a unique opportunity to study the spectral evolution of the source along with the temporal one for Neutron star systems \citep{klitzing:2020ATel13538....1C, klitzing:2017ATel10452....1R, klitzing:2017ApJ...841...41V, klitzing:2018ApJ...860...88B}. Such studies have also been undertaken for Cyg X-2 \citep{klitzing:Devasia_2021,klitzing:10.1093/mnrasl/slac014} and the Cyg-like Z source GX 5-1 \citep{klitzing:2019RAA....19..114B}. From these recent analyses and studies, a number of interesting facts about these Cyg X-2 or Cyg-like sources have come into sight. Detection of 42 Hz QPO along with a broadband noise around 10 Hz in the power density spectra (PDS) of Cyg X-2 with the changing sign of time lag reveals that the hard lags occur when coronal variation is delayed compared to that of soft photon source
and the soft lag occurs when the soft photon source variation delayed to the coronal variation. On the other hand, spectral analysis of GX 5-1 reveals the existence of the accretion disk and the accretion disk flux ratio is the primary driver of the different positions in the HID diagram whereas the temporal analysis reveals the existence of the QPO near 50 Hz which changes as the source moves from HB to NB and energy-dependent r.m.s variations suggest that QPO has originated from the coronal region.\\
In this work, we report the spectral and temporal analysis of GX 340+0 with {\it AstroSat} in order to characterize the spectral evolution and to study the energy-dependent properties of the QPOs and broadband noise.

\section{Observation \& Data Reduction}
GX 340+0 was observed using {\it AstroSat} payloads SXT and LAXPC from 30th July 2017  to 31st July 2017, for a total exposure of 40 ks with observation ID [G07\_016T01\_9000001420] spanning from orbits  09939 to 09959 or 16 orbits. The observational data is publicly available from {\it AstroSat} Archive\footnote{\url{https://astrobrowse.issdc.gov.in/astro\_archive/archive/Home.jsp}}. The data were reduced using the standard software LAXPCSoftware\_Aug4 available on the ASSC page\footnote{\url{https://www.tifr.res.in/~astrosat\_laxpc/LaxpcSoft.html}}
\footnote{\url{https://www.tifr.res.in/~astrosat\_sxt/sxtpipeline/AS1SXTLevel2-1.4a\_Standard\_Operating\_Procedure\_V1.4a.pdf}}.  All orbits level 1 event files were merged to a single level 2 event file and a good time interval (gti) was created by removing earth occultation and SAA period.  The standard software was used to generate light curve, background light curve, energy spectra, power density spectra (PDS), r.m.s, and time lag [{http://astrosat-ssc.iucaa.in/uploads/threadsPageNew\_LAXPC.html}]. For the timing analysis, all three LAXPC components were used in the energy range of 3-20 keV. For the photon spectra,  LAXPC 20 was used and a 3\% systematic error was included in all analyses.

The SXT level-1 data were processed using SXTPIPELINE to generate the level-2 data \footnote{ \url{http://astrosat-ssc.iucaa.in/laxpcData}}.  The average count rates are found well below 40 counts/sec, and hence pileup correction was not done.  We selected a 626 arc-sec region around the central part and extracted the lightcurve,  spectra using the {\bf XSELECT} V2.4k. The energy spectrum of LAXPC and SXT were fitted jointly using  {\bf Xspec} 12.11.1 \citep{klitzing:1996ASPC..101...17A}.\\

\section{Lightcurves \& Hardness Ratios }
\begin{figure}
	\includegraphics[width=1.0\linewidth]{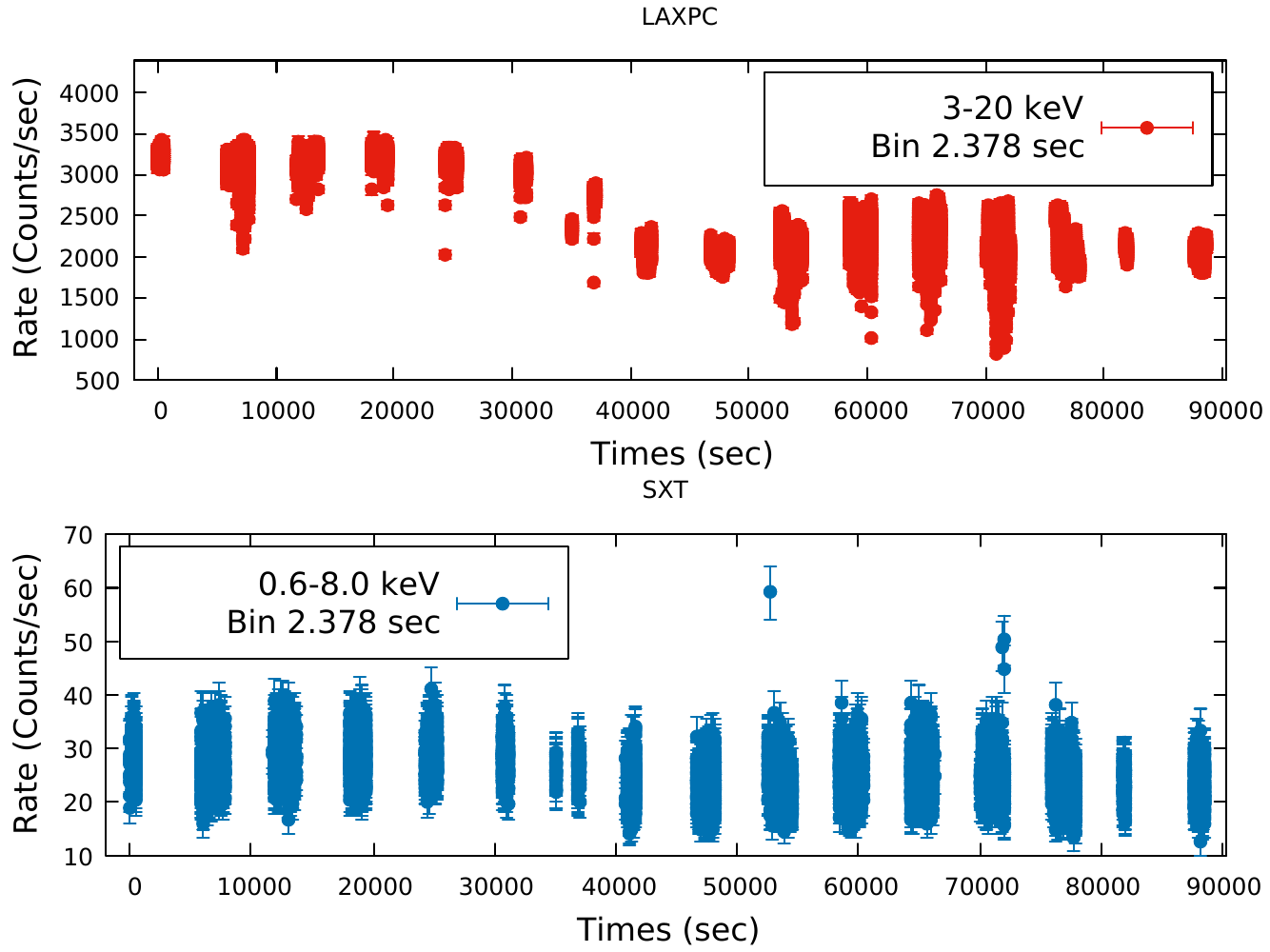}
	\caption{Background subtracted light curve of GX 340+0 using time bin 2.3778 seconds at the SXT time resolution.LAXPC light curve is in the energy range 3-20 keV using LAXPC 10, 20 and the SXT light curve is in the energy range 0.6-8 keV.}
	\label{Fig1}
\end{figure}
Figure \ref{Fig1} shows the background subtracted light curve of LAXPCs and SXT at 3-20 keV and 0.6-8 keV respectively using the SXT time resolution of 2.378 seconds. In the upper panel, the light curve is obtained using LAXPC counters 10, 20 and SXT at the bottom panel. The SXT exposure time is less compared to LAXPC because SXT operates only at night time. For the energy spectra, only strictly simultaneous data from LAXPC and SXT have been used.

\begin{figure}	
  	\includegraphics[width=1.17\linewidth]{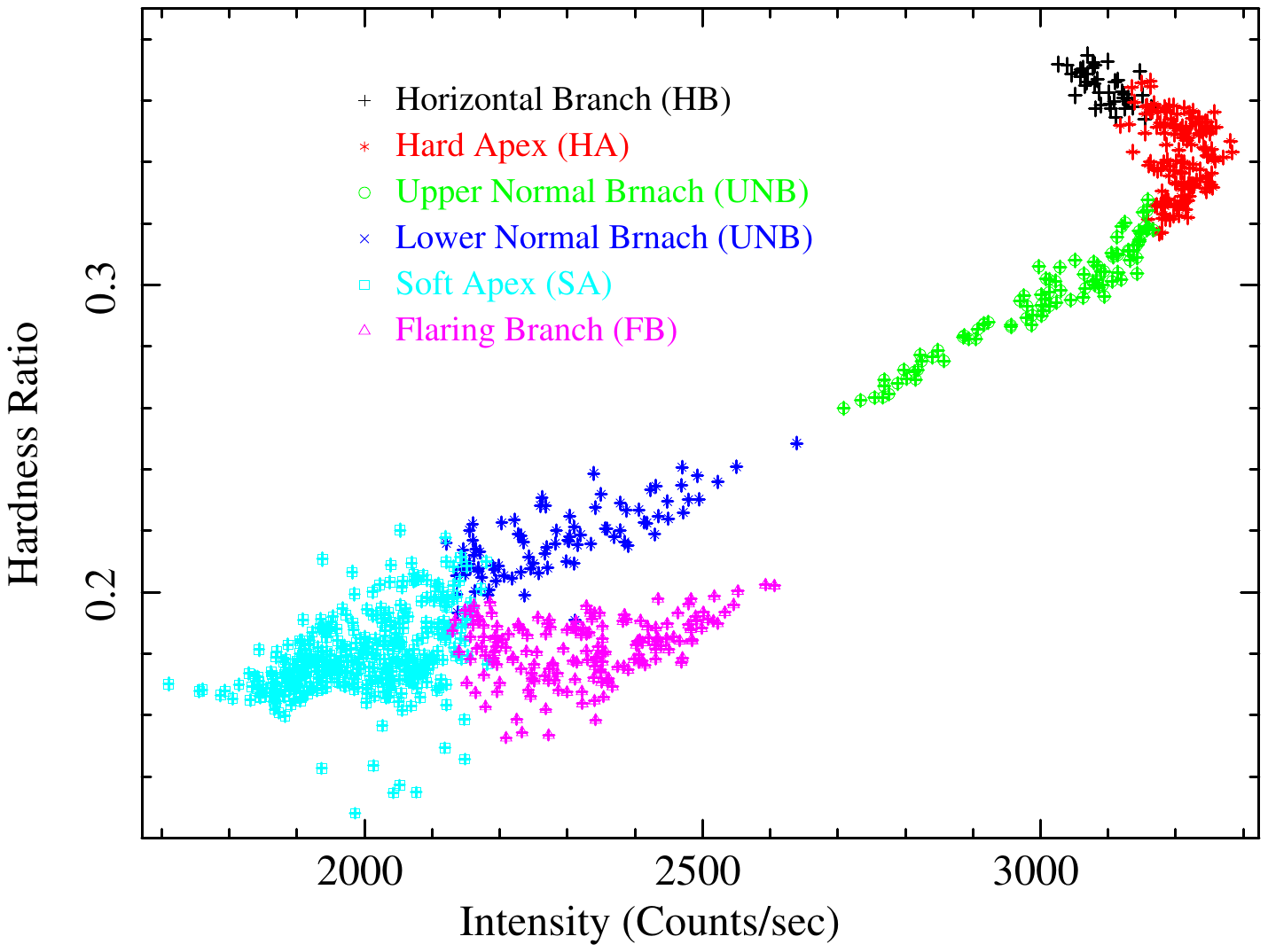}
	\caption{Hardness Intensity Diagram (HID) of GX 340+0 observed using LAXPC 10 and 20 with time bin 40 sec within the energy range of 3-20 keV using 8-20 keV as hard energy band and 3-8 keV as soft one. The total HID is divided into six segments which are shown in the figure.}
	\label{Fig2}
\end{figure}

Figure \ref{Fig2} shows the Hardness-Intensity-Diagram (HID) by using LAXPC units 10 and 20. The HID shows the ratio of the intensity of hard (8-20 keV) to soft energy (3-8 keV) with respect to total intensity. The hardness ratio plot was binned at  40 seconds. In the HID, all Z-track branches (short HB, NB, and FB) are evident. The short HB corresponds to the beginning of the data in lightcurve \ref{Fig1} and with time the source moved down to FB through NB tracing the whole diagram.  We segregate the HID into 6 different portions to understand the timing and spectral evolution along Z-track as shown in \ref{Fig2}. The split segments' intensity, ratio, and exposure time are tabulated in Table \ref{Table1}.
\begin{table}
	\centering
	\begin{tabular}{|c|c|c|c|c|}
		\hline
	{\bf Loc} & {\bf Count-rates} & {\bf Hardness-ratio} & {\bf Exp. Time(sec)} & {\bf LAXPC} \\
	& & & [simultaneous] & [Sec]\\
	\hline
	HB & 2900-3100 & 0.39-0.41 & 518 & 1600 \\
	HA & 3100-3300 & 0.35-0.39 & 4270 & 6000\\
	UNB & 2700-3100 & 0.30-0.35 & 1002 & 3764\\
	LNB & 2150-2650 & 0.25-0.30 & 863 & 3480\\
	SA & 1700-2150 & 0.20-0.25 & 6123 & 15760 \\
	FB & 2150-2600 & 0.19-0.23 & 3419 & 6640\\	
	\hline 
	\end{tabular}
	\caption{Location of the different segments in the hardness-ratio diagram based on count rates and the hardness-ratio. Simultaneous LAXPC 20 and SXT exposure time (selected region) of observations in 6 segments has been shown in the third column for spectral analysis. The last column represents the LAXPC exposure times of the corresponding segments.}
	\label{Table1}
\end{table} 
\begin{table*}
	\centering
	
	\begin{tabular}{|l|c|c|c|c|c|c|c|r|}
		\hline
		\hline
		{\bf Parameters} &  {\bf HB}  &  {\bf HA}  &  {\bf UNB}  & {\bf LNB}  &  {\bf SA}  &  {\bf FB}  \\
		\hline
		{{$\bf N_{H}$}} & $4.61\pm0.08$ & $4.38\pm0.10$ & $4.49\pm0.16$ & $4.35\pm0.17$ & $4.25\pm0.08$ & $4.11\pm0.10$ \\
		\hline
		{$\bf kT_{bb}$} & $1.25\pm0.04$ & $1.26\pm0.02$ & $1.22\pm0.03$ & $1.13\pm0.03$ & $1.15\pm0.01$ & $1.27\pm0.01$\\
		\hline
		{\bf Bbody-norm} & $481.6\pm87.9$ & $511.9\pm45.6$ & $634.4\pm71.2$ & $616.1\pm69.3$ & $519.9\pm31.2$ & $432.1\pm22.2$ \\
		\hline
		{\bf Radius(km)} & $24.1\pm2.1$ & $24.9\pm1.9$ & $27.7\pm1.5$ & $27.3\pm1.5$ & $25.1\pm0.7$ & $22.9\pm0.6$ \\
		\hline
		{\bf tau ($\tau_{opt}$)} & $10^f$ & $10^f$ & $10^f$ & $10^f$ & $10^f$ & $10^f$ \\
		\hline
		{ $\bf kT_e$} & $3.51\pm0.11$  & $3.40\pm0.07$ & $3.33\pm0.07$ & $3.02\pm0.02$ & $3.07\pm0.10$ & $3.70\pm0.37$ \\
		\hline
		{\bf cov\_frac} & $0.43\pm0.04$ & $0.39\pm0.03$ & $0.36\pm0.03$ & $0.24\pm0.03$ & $0.12\pm0.01$ & $0.04\pm0.01$\\
		\hline
		{\bf blackbody Flux} & $1.27\pm0.04$ & $1.52\pm0.02$ & $1.54\pm0.04$ & $1.11\pm0.03$ & $1.05\pm0.01$ & $1.24\pm0.03$ \\
		\hline
		{\bf Bolometric Flux} & $1.58\pm0.04$ & $1.82\pm0.03$ & $1.81\pm0.04$ & $1.23\pm0.02$ &  $1.12\pm0.02$ & $1.26\pm0.02$ \\
		\hline
		{\bf Compton Flux} & $0.32\pm0.02$ & $0.29\pm0.01$ & $0.27\pm0.01$ & $0.12\pm0.01$ & $0.07\pm0.01$&$0.03\pm0.01$ \\
		\hline
		{\bf Blackbody Flux Ratio} &  $0.80\pm0.003$ & $0.84\pm0.01$ & $0.85\pm0.01$ & $0.90\pm0.01$ & $0.94\pm0.01$ & $0.98\pm0.01$ \\
		\hline
		{\bf Compton Flux Ratio} & $0.20\pm0.004$ & $0.16\pm0.001$ & $0.15\pm0.03$ &  $0.09\pm0.01$&  $0.06\pm0.01$& $0.02\pm0.001$ \\
		\hline
		{\bf Chisq/dof} & 83.57/90 & 137.01/104 & 111.51/97 & 105.58/91 & 98.27/105 & 126.22/100 \\
		\hline
		\hline
	\end{tabular}	
	\caption{This table presents the best-fitted parameters obtained from the spectral studies using the model ({\tt const}*{\tt tbabs}*{\tt thcomp}*{\tt bbodyrad}). $N_H$ is the absorption column density and is computed in the unit of $10^{22}$ atoms cm$^{-2}$. Tau is the Thomson-optical depth of the comptonized component and $kT_e$ represents the temperature of the comptonized electron which is calculated in keV and cov\_frac is the covering fraction. $kT_{bb}$ is the blackbody temperature in keV. Norm is representing the blackbody norm and is related to the radius of the neutron star. $(^f)$ represents the frozen parameter. Bolometric flux is in the unit of ergs cm$^{-2}$s$^{-1}$. The Compton flux ratio is defined as the ratio of Compton flux to the total flux and the blackbody flux ratio is defined to be the ratio of blackbody flux to the total unabsorbed bolometric flux.}
	\label{Table2}
\end{table*}
\begin{figure}
	\includegraphics[width=1.0\linewidth]{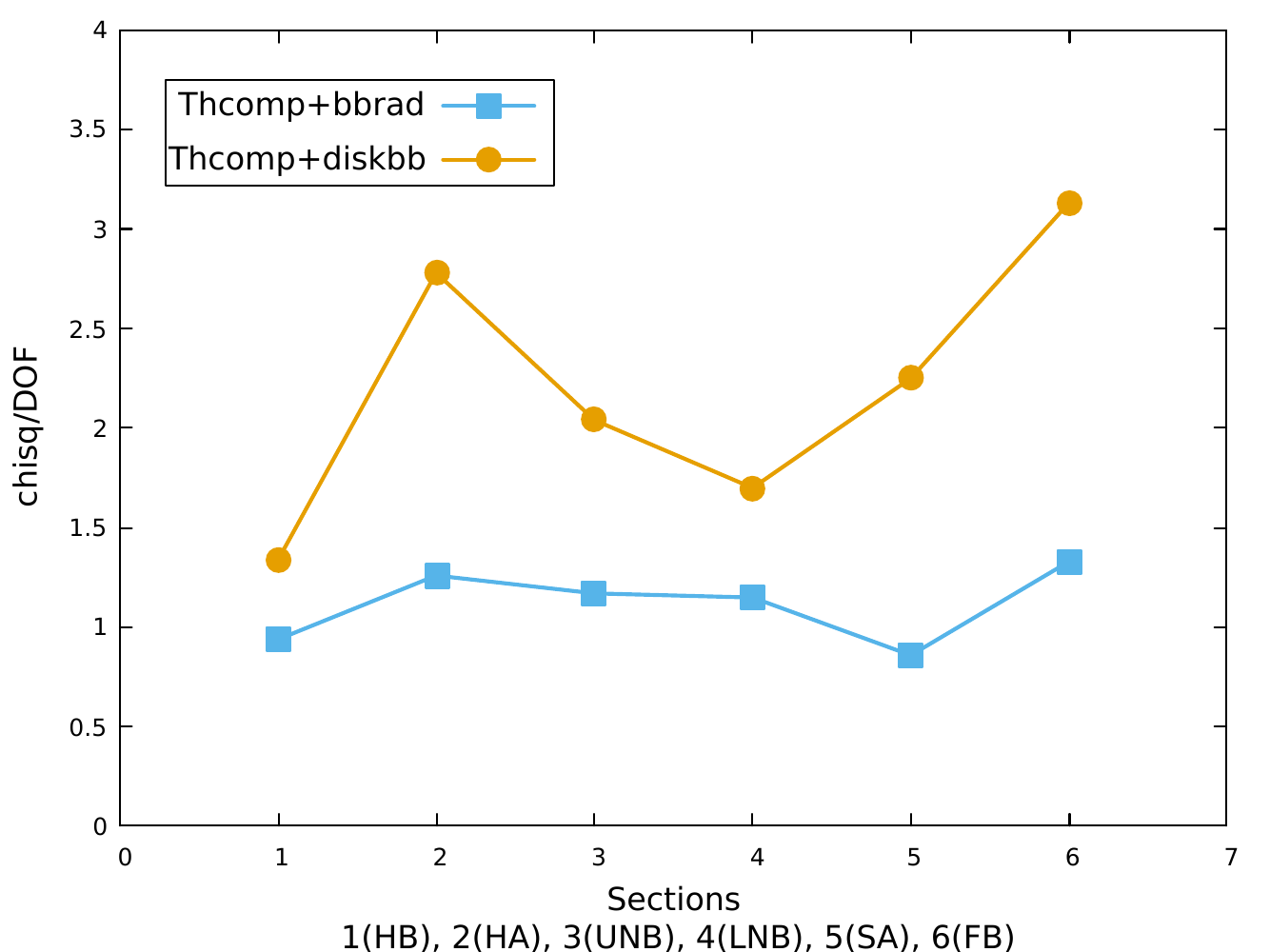}
	\caption{Comparison of reduced $\chi^2_{red}$ of the two models that have been used to fit the observed spectra.}
	\label{Fig3}
\end{figure}
\begin{figure}
	\centering
	\includegraphics[width=1.17\linewidth]{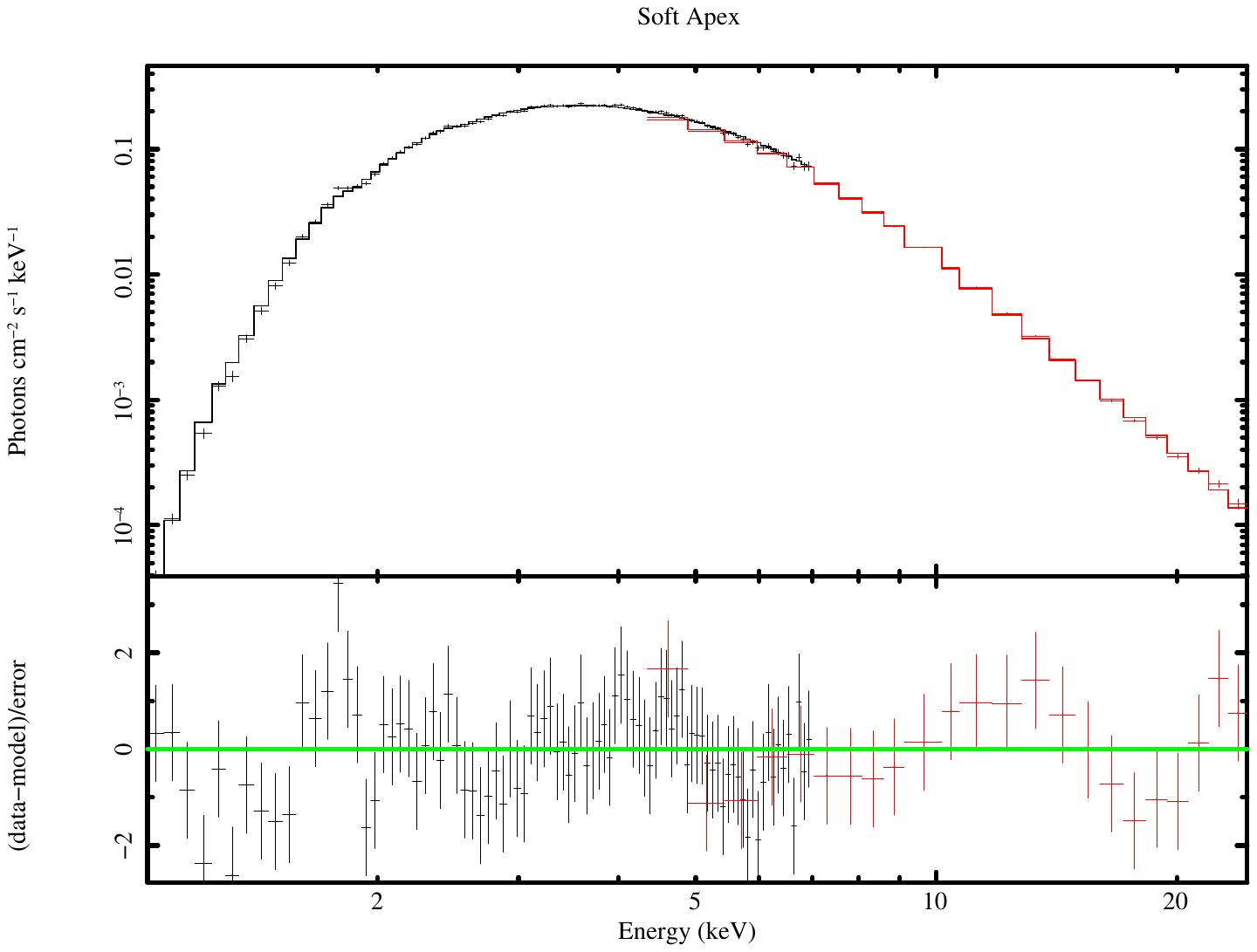}
	\caption{Photon spectrum of GX 340+0 in the Soft Apex branch using both sxt and laxpc in the energy range 1-25 keV.The model that has been used is {\tt const*tbabs*(thcomp*bbodyrad)} }
	\label{Fig4}
\end{figure}	
\begin{figure*}
	\centering
	\begin{tabular}{lcccr}
		\includegraphics[width=0.45\linewidth]{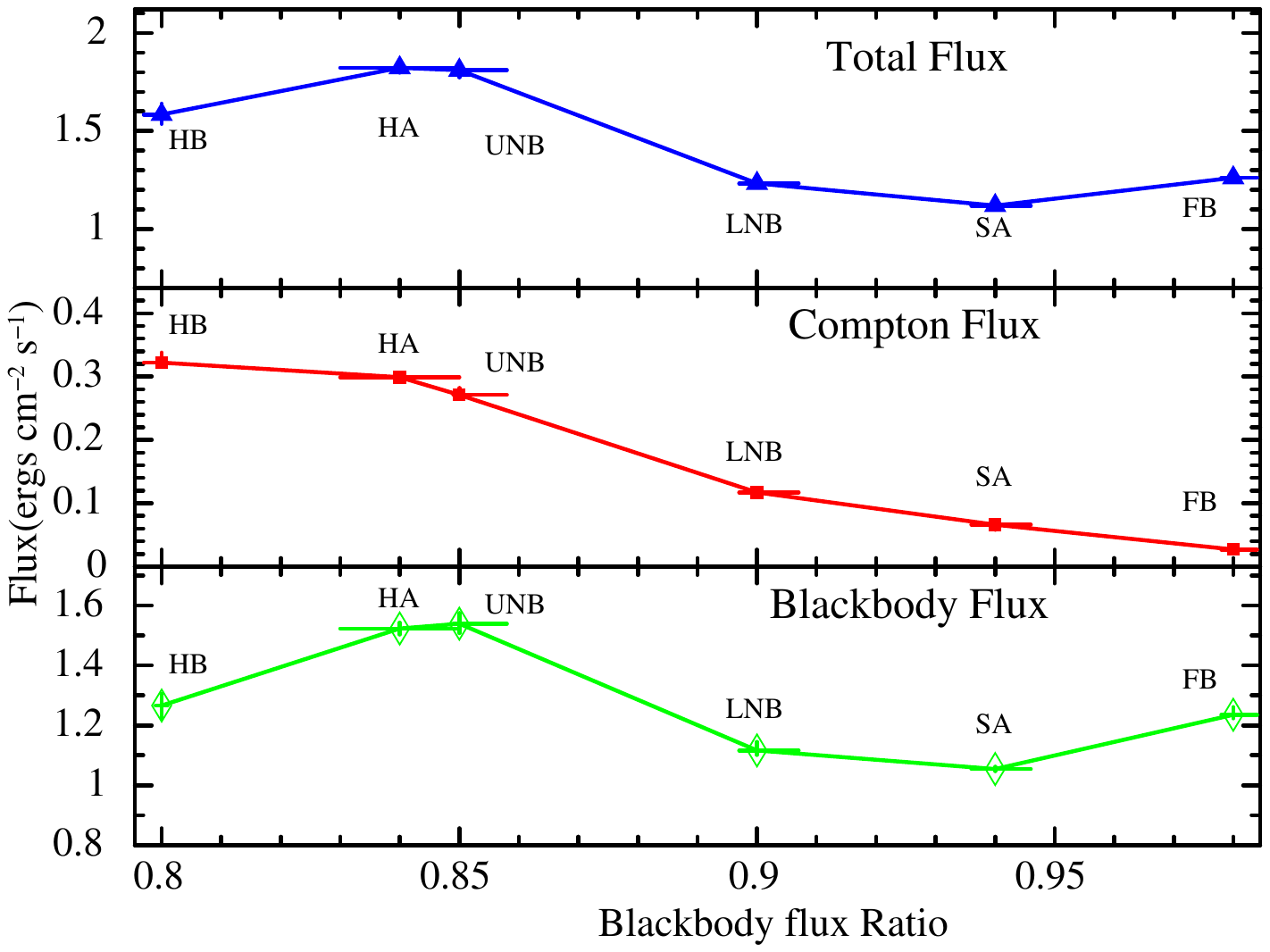} & \includegraphics[width=0.45\linewidth]{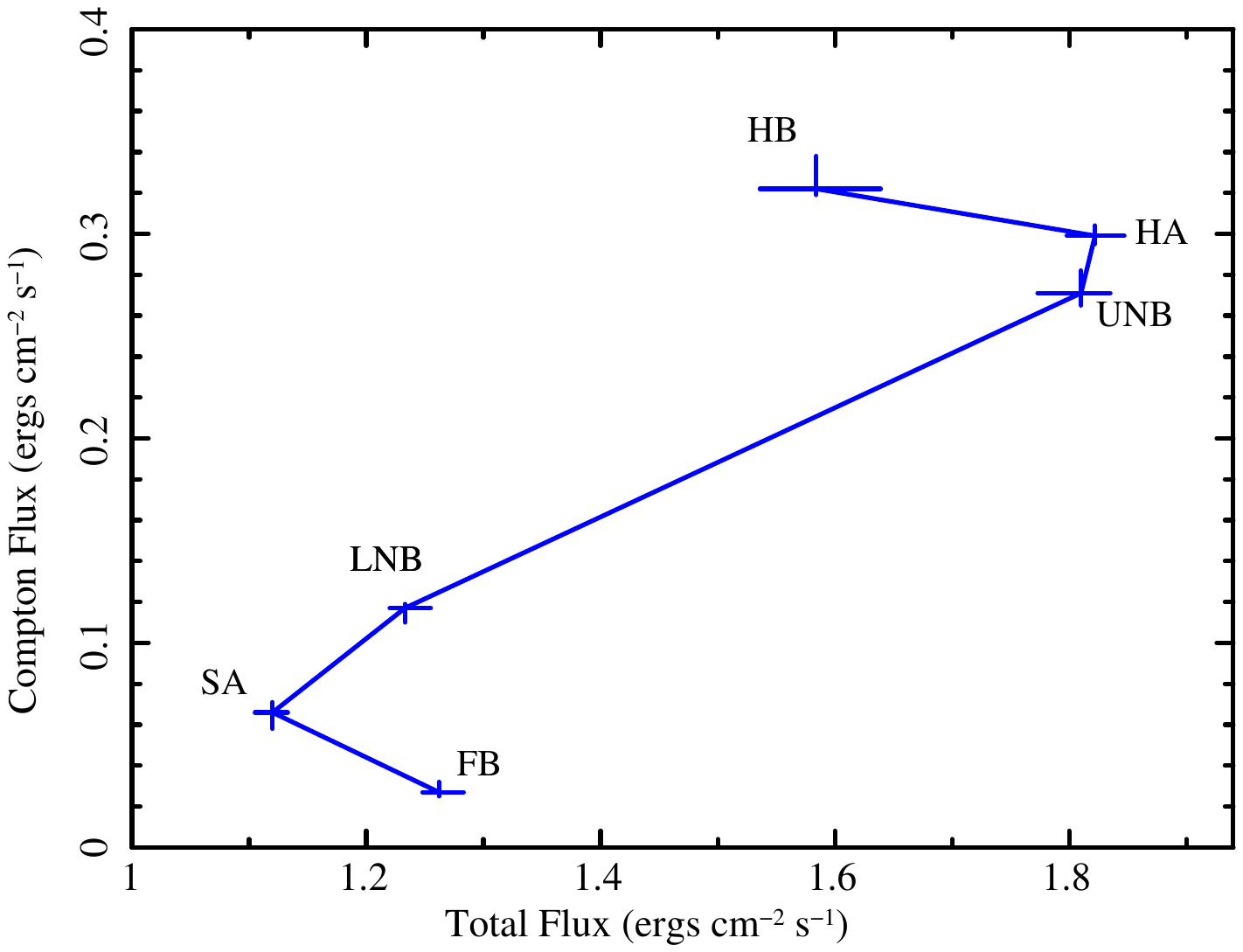} \\
		\includegraphics[width=0.45\linewidth]{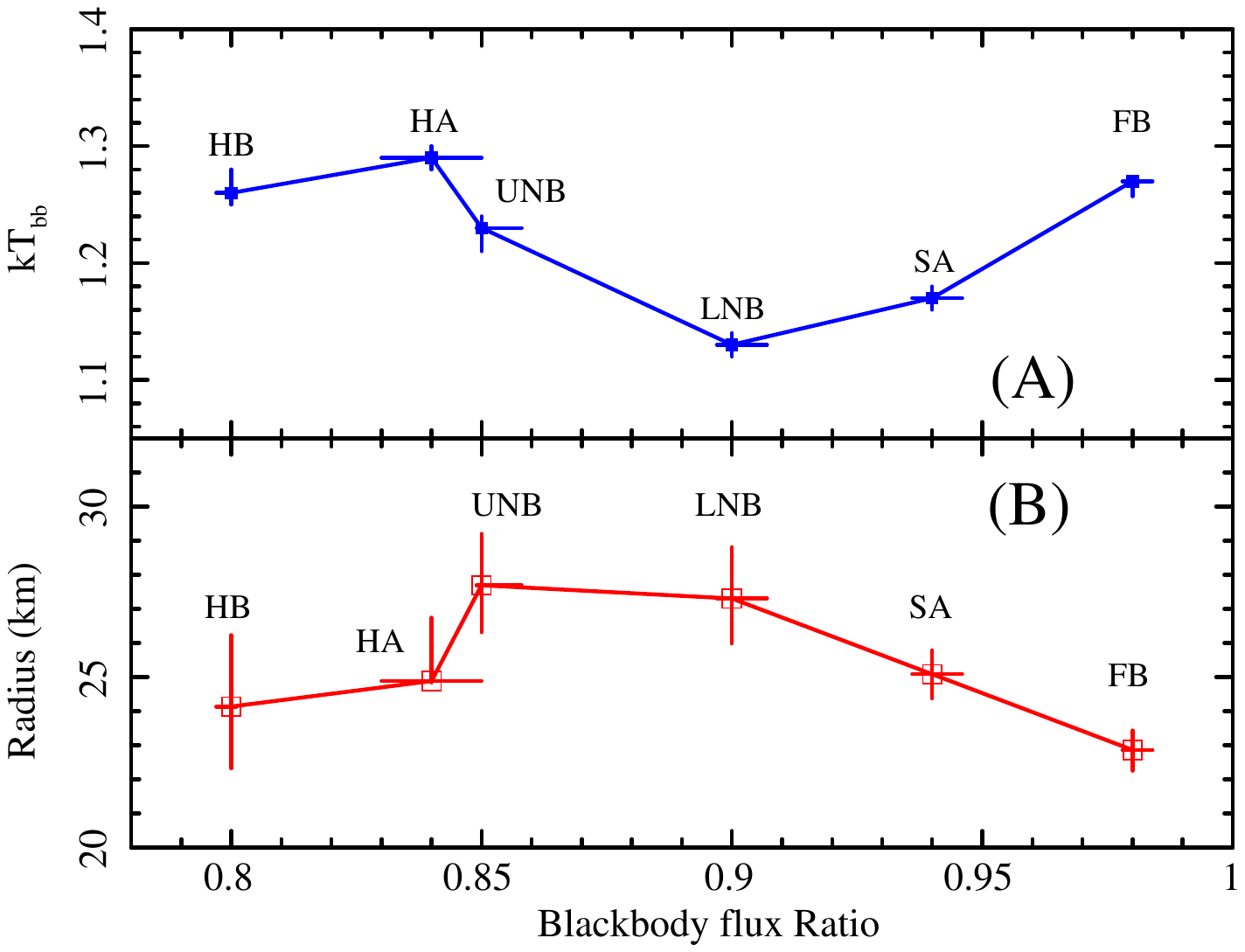}&
		\includegraphics[width=0.45\linewidth]{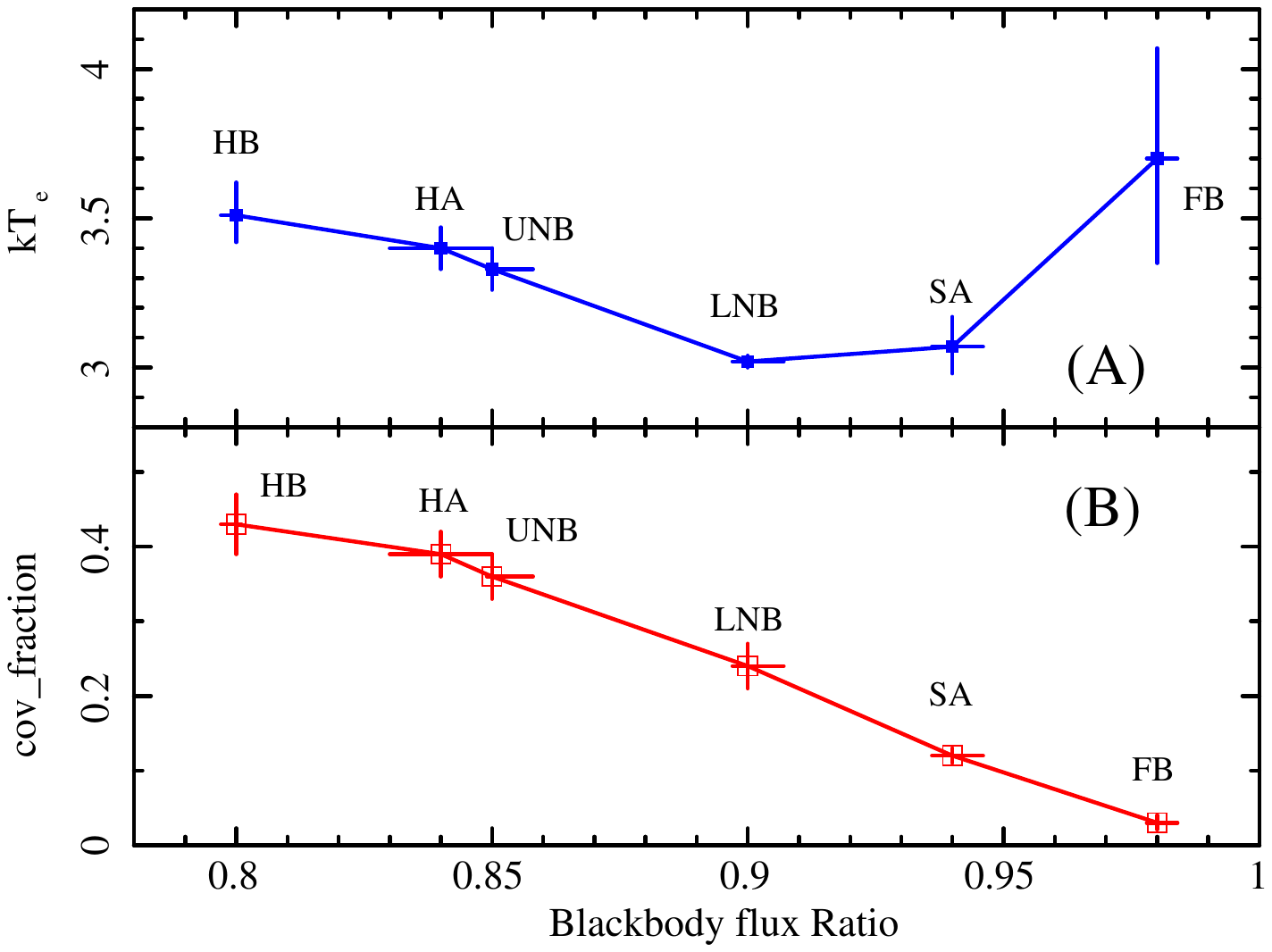}\\
	\end{tabular}
	\caption{{\bf Left Upper}  The plot of the variation of the blackbody flux and total unabsorbed flux and Compton flux versus blackbody flux ratio. {\bf Right Upper}  Plot of variation of Compton flux with respect to the Total unabsorbed flux.\\{\bf Bottom left} (A) shows the variation of blackbody temperature with respect to the blackbody flux ratio and (B) shows the variation of radius (km) with respect to the variation of blackbody flux ratio. {\bf Bottom right} (A) shows the variation of electron temperature with respect to the blackbody flux ratio and (B) shows the variation of the covering fraction with respect to the disk flux ratio. }
	\label{Fig5}
\end{figure*}
\begin{figure*}
	\centering
	\begin{tabular}{lcccr}
		\includegraphics[width=0.45\linewidth]{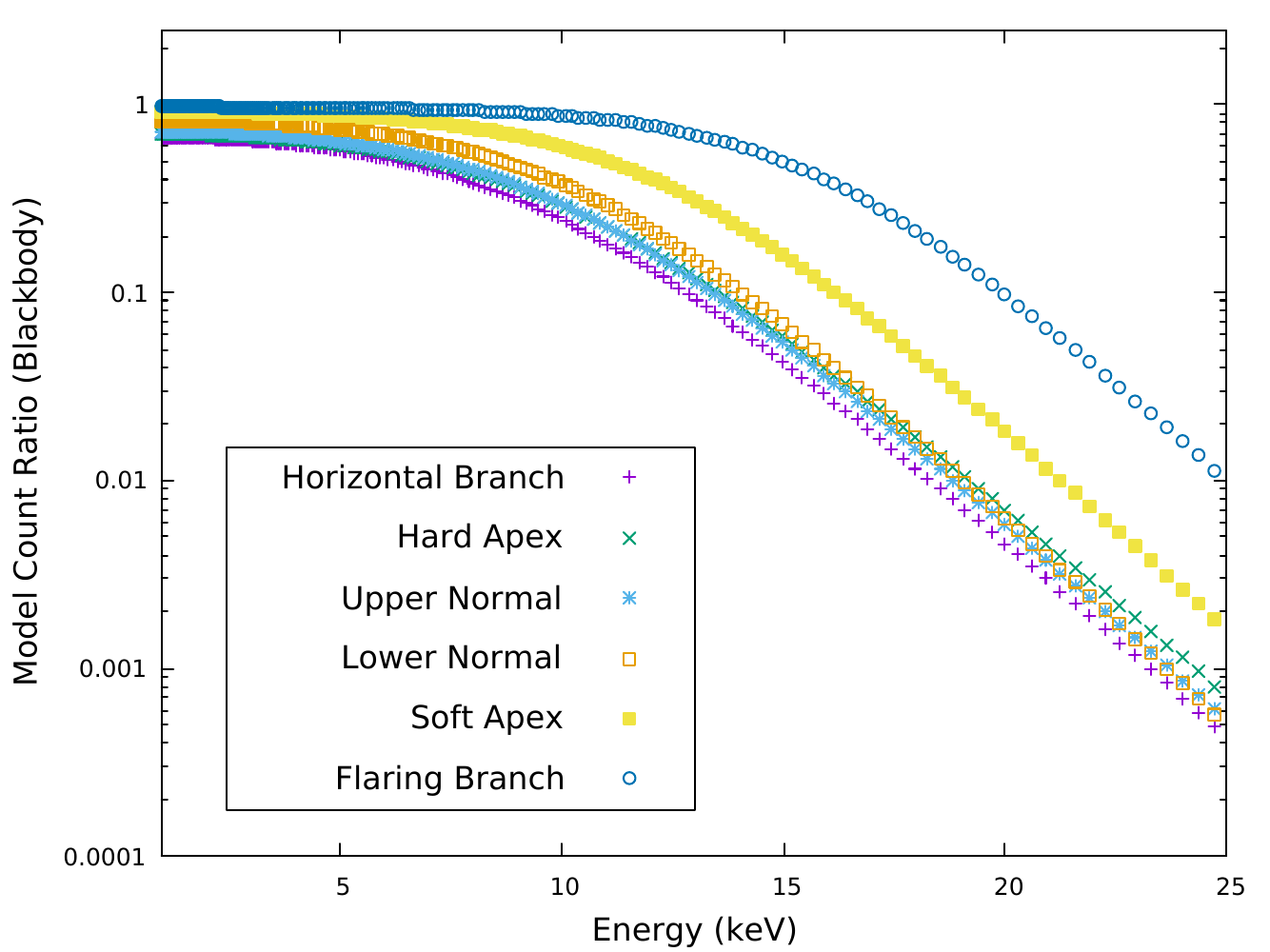} & \includegraphics[width=0.45\linewidth]{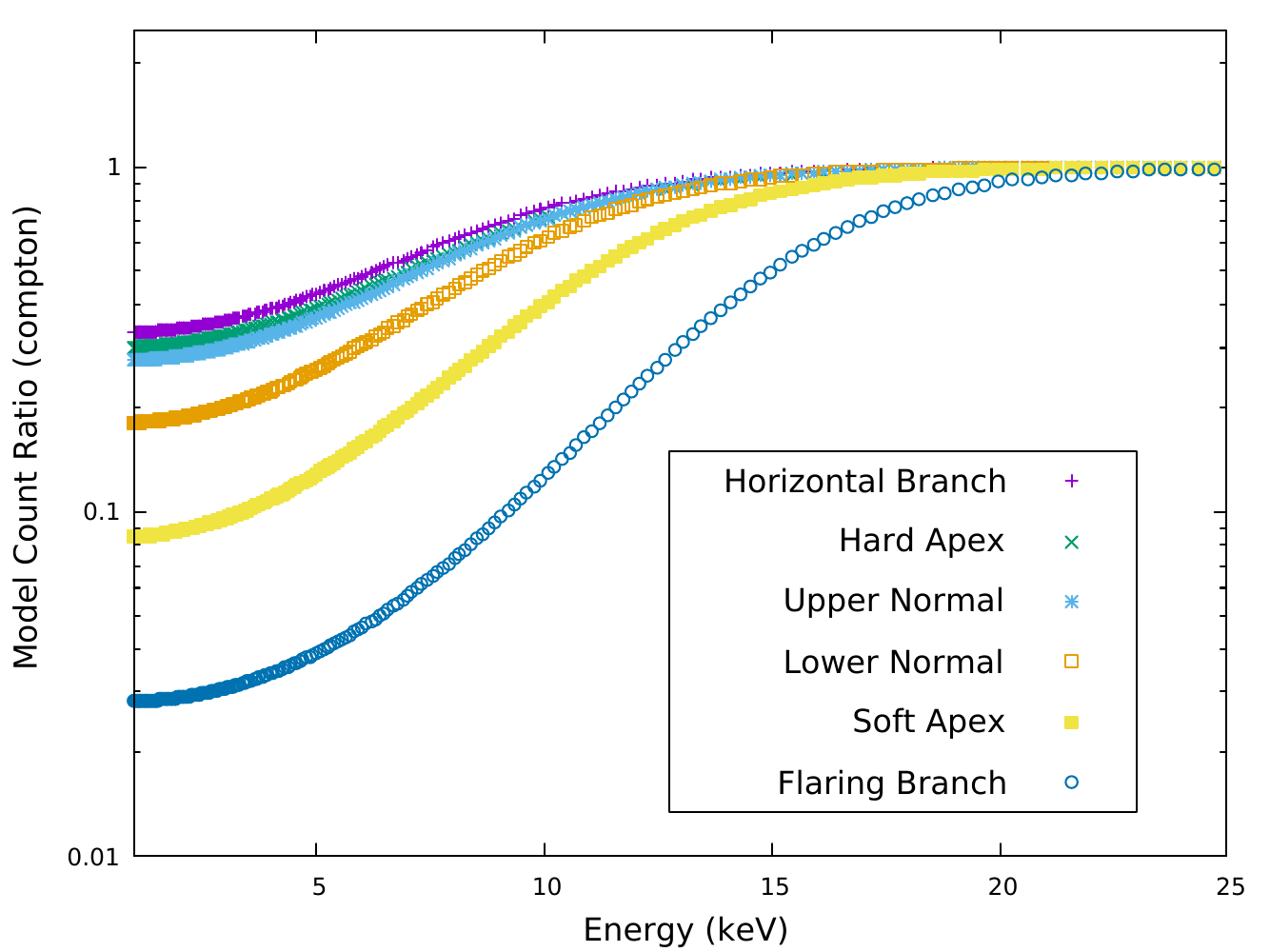} \\
	\end{tabular}
	\caption{Plots show the variation of the model count ratio with respect to the energy in different branches. The model count ratio is defined as the ratio of counts obtained using (1-f) times the counts ratio of {\tt const*tbabs*bbodyrad} to {\tt const*tbabs*(thcomp*bbodyrad)}. f is the best fit covering fraction obtained from the fitting. The left shows the blackbody count ratio variation whereas the right one shows the Compton count ratio variation with respect to energy. The Compton counts have been obtained by subtracting the counts of blackbody counts ratio from 1. }
	\label{Fig6}
\end{figure*}
\begin{table*}
	\centering
	\begin{tabular}{lccccccccr}
		\hline
		\hline
		{\bf Parameters} & {\bf HB} & {\bf HA} & {\bf UNB } & {\bf LNB} & {\bf SA } & {\bf FB} \\
		\hline
		$\nu_0$ (Hz) & $41.3\pm0.4$ & $45.5\pm 0.7$ & $50.5\pm 1.5$ & $6.2\pm 0.3$ & $5.9\pm 0.1$  & $6.5\pm 0.2$  \\
		$\sigma_0$ (Hz) & $7.9\pm 1.5$  & $15.1\pm4.9$ & $11.1\pm7.0$ & $2.1\pm1.2$ & $2.1\pm0.4$ & $1.3\pm0.5$ \\
		$N_0\times 10^{-4}$ & $13.7\pm1.8$ & $8.1\pm3.0$ & $3.6\pm1.4$ & $2.7\pm1.3$ & $4.3\pm0.9$ & $1.7\pm0.4$ \\
		\hline
		$\nu_1$ (Hz)& $1.1\pm1.2$  & $6.3\pm2.8$ & $0.9\pm1.2$& $1.5\pm0.1$ & $2.9\pm0.1$ & $1.9\pm0.1$  \\
		$\sigma_1$ (Hz)& $7.9\pm3.9$ & $19.4\pm3.4$ & $12.6\pm1.8$& $1.5\pm 0.2$ & $0.7\pm0.2$ & $1.3\pm0.1$  \\
		$N_1\times 10^{-4}$ & $10.3\pm3.2$ & $7.2\pm1.2$ &$ 10.4\pm0.9$ & $21.7\pm2.2$ & $4.1\pm1.0$& $16.3\pm1.9$ \\
		\hline
		$\nu_2$ (Hz)& $18.2\pm3.5$ & - & - & - & $10.4\pm0.9$ & -  \\
		$\sigma_2$ (Hz) & $19.0\pm18.5$ & - & - & -  & $4.4\pm4.2$ & - \\
		$N_2\times 10^{-4}$ & $7.5\pm3.7$ & - & - & - &  $2.5\pm1.6$ & -  \\
		\hline
		\hline
	\end{tabular}
	\caption{Details of best-fitted parameters of the Lorentzian fitted for the QPOs found in the different segments.Here $\nu_0 ,\nu_1,\nu_2 $ is representing frequency in Hz, $\sigma_0,\sigma_1,\sigma_2$ is representing width, $ N_0,N_1,N_2$ is representing normalization of Lorentzian . }
	\label{Table3}	
\end{table*}
\begin{table}
	\centering
	\begin{tabular}{|c|c|c|}
		\hline
		{\bf Branch} & {\bf Frequency (Hz)} & {\bf $\Delta \nu$}\\ 
		\hline
		{\bf HB} & 30.04 Hz & 7.82 Hz \\
		\hline
		{\bf HA} & 41.25 Hz & 13.72 Hz \\
		\hline
		{\bf UNB} & 54.68 Hz & 9.01 Hz \\
		\hline
		{\bf LNB} & 5.00 Hz & 2.50 Hz \\
		\hline
		{\bf SA} & 15.62 Hz & 3.90 Hz \\
		\hline
		{\bf FB} & 6.25 Hz & 1.25 Hz \\
		\hline
		
	\end{tabular}
	\caption {  Frequency (Hz) is the frequency at which the lags and r.m.s are calculated, {\bf $\Delta \nu$} is the frequency resolution for each of the case.}
	\label{Table4}
\end{table}

\begin{figure*}
	\centering
	
	\begin{tabular} {l c c r}
		\includegraphics[width=0.452\linewidth]{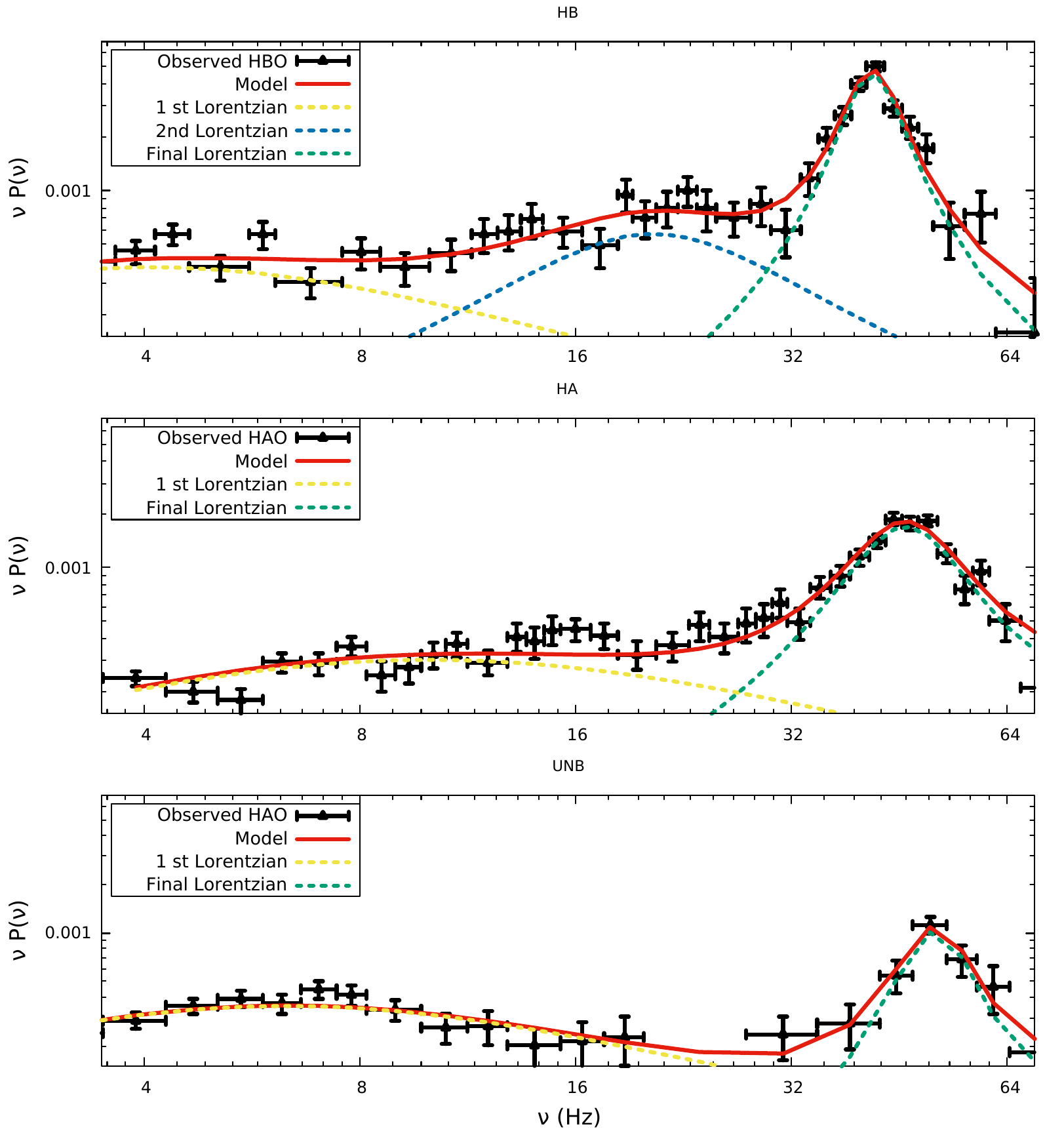}&\includegraphics[width=0.452\linewidth]{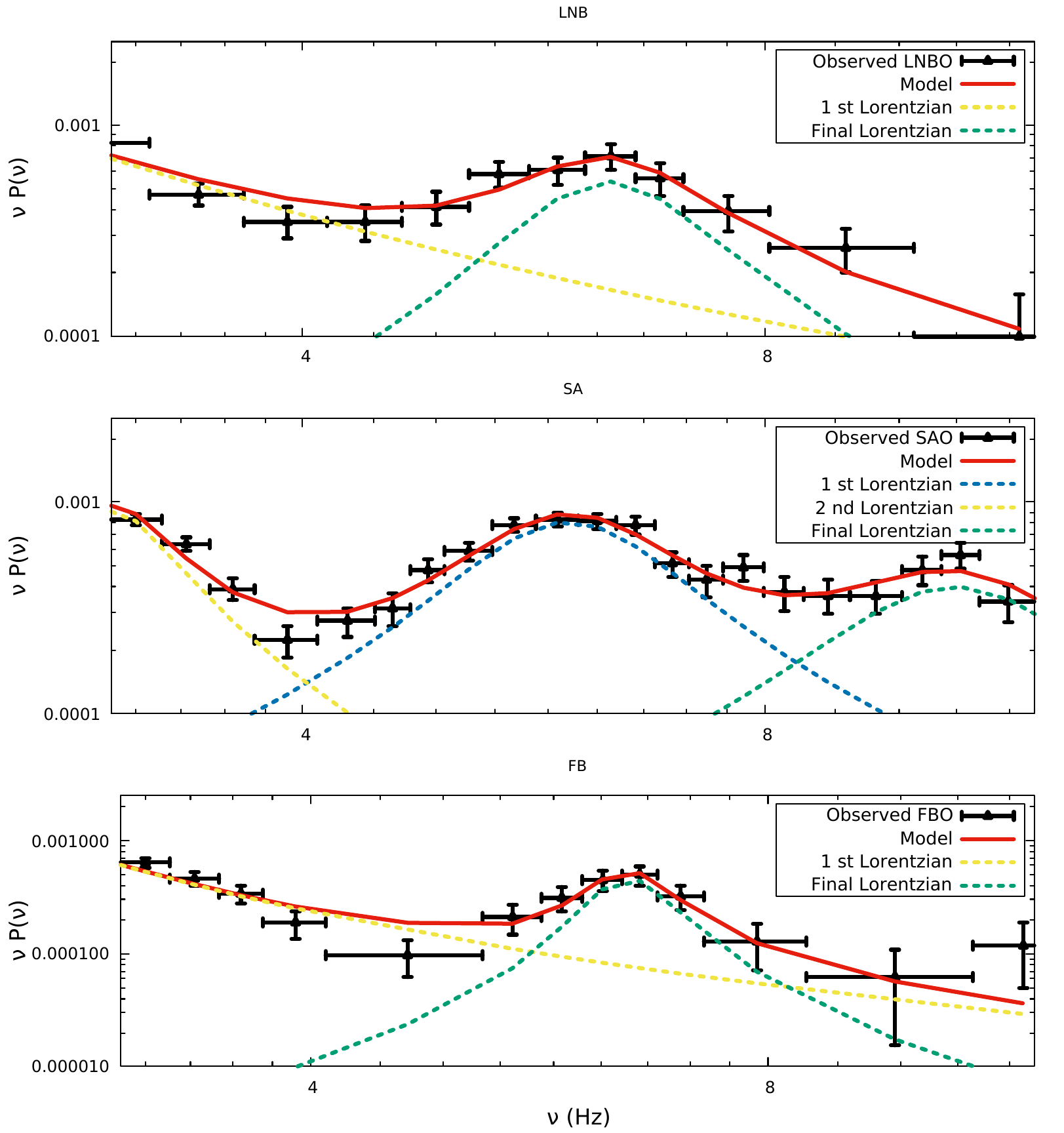}
	\end{tabular}
    \caption{$AstroSat$ LAXPC power density spectra of GX 340+0 in the energy range 3-20keV using all LAXPC units. Power density spectra have been fitted using Lorentzian and the best-fitted values are listed in Table \ref{Table3}. $\nu$ represents the frequency and P($\nu$) represents the power.}
    \label{Fig7}
	
\end{figure*}
\begin{figure*}
	\centering
	\begin{tabular}{lcr}
		\includegraphics[width=0.32\linewidth]{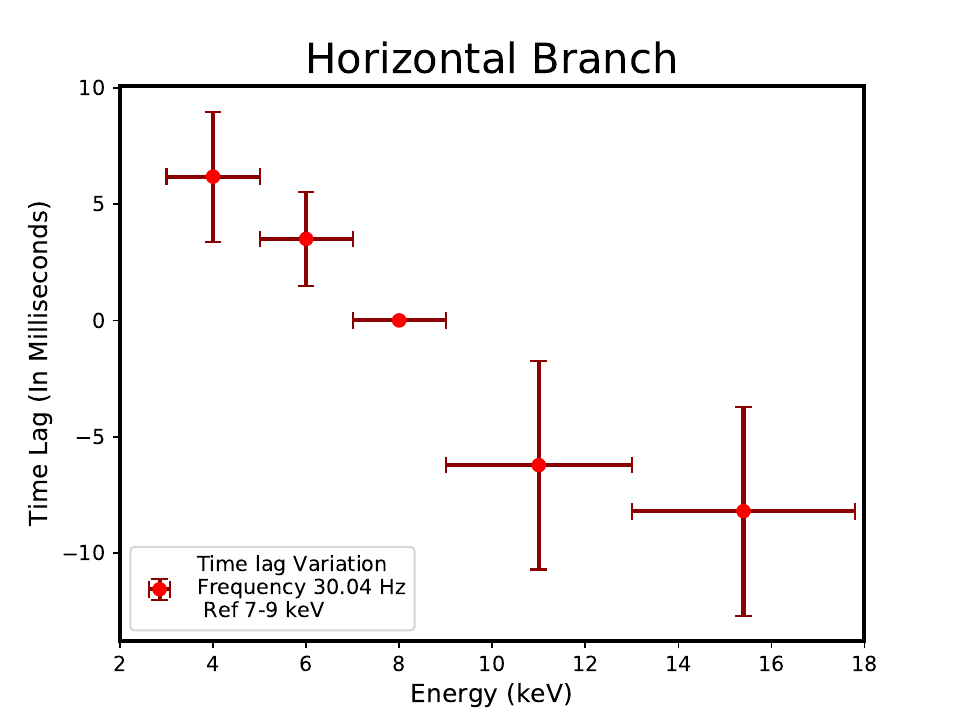} & \includegraphics[width=0.32\linewidth]{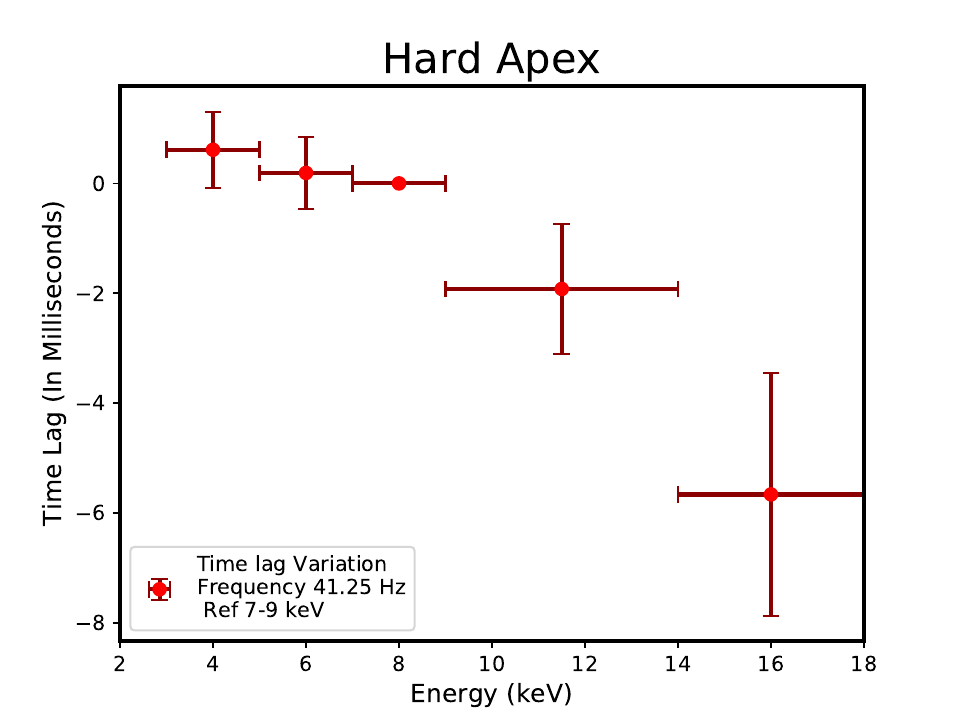} & \includegraphics[width=0.32\linewidth]{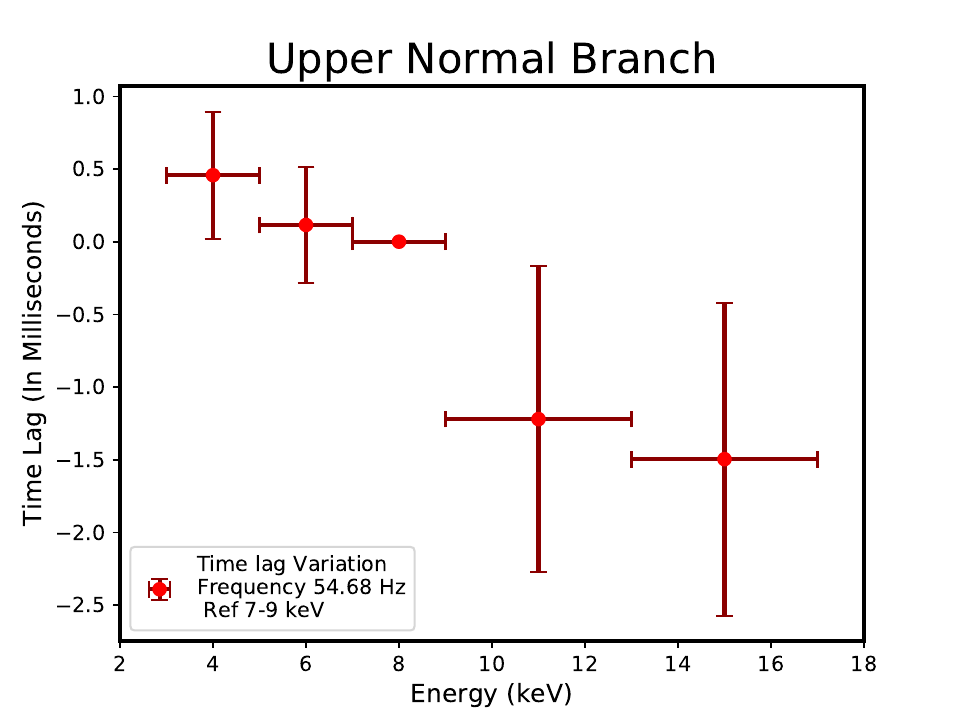}\\
		\includegraphics[width=0.32\linewidth]{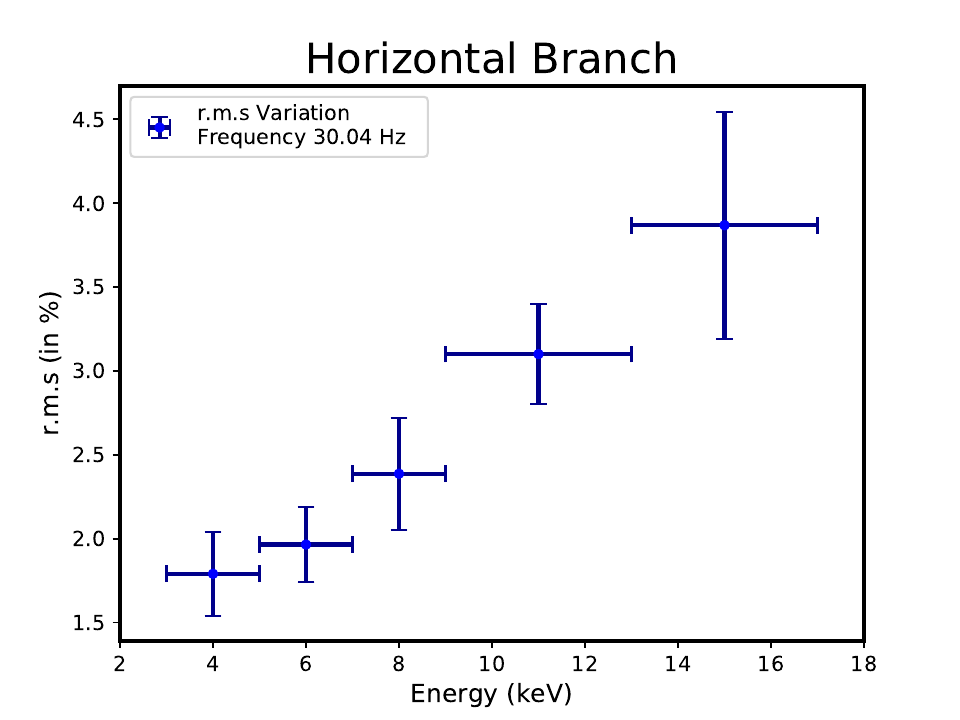} &
		\includegraphics[width=0.32\linewidth]{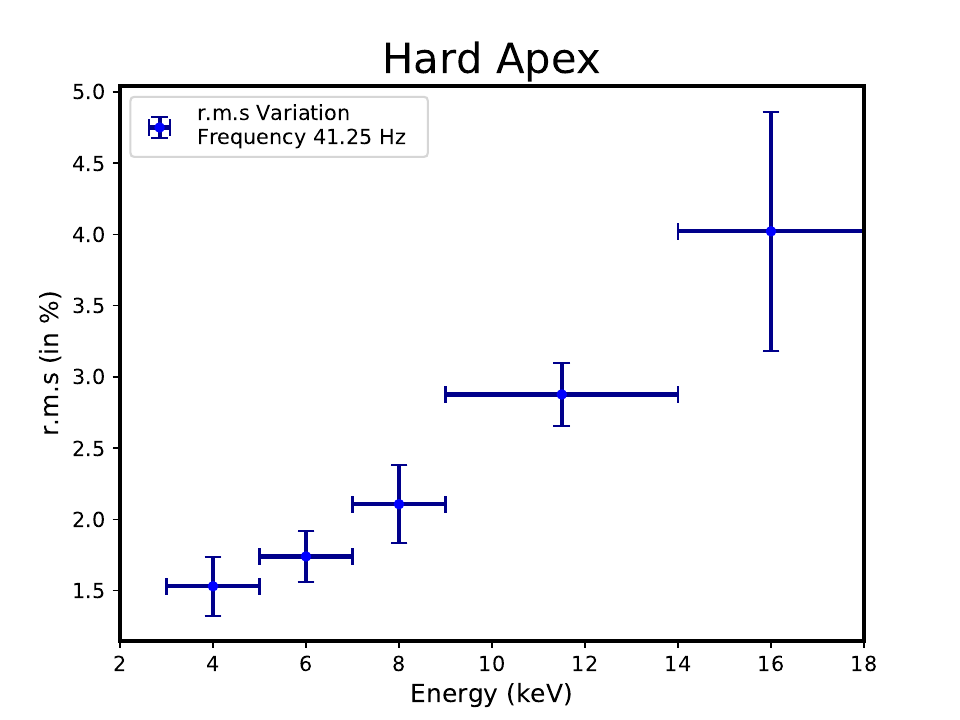} &
		\includegraphics[width=0.32\linewidth]{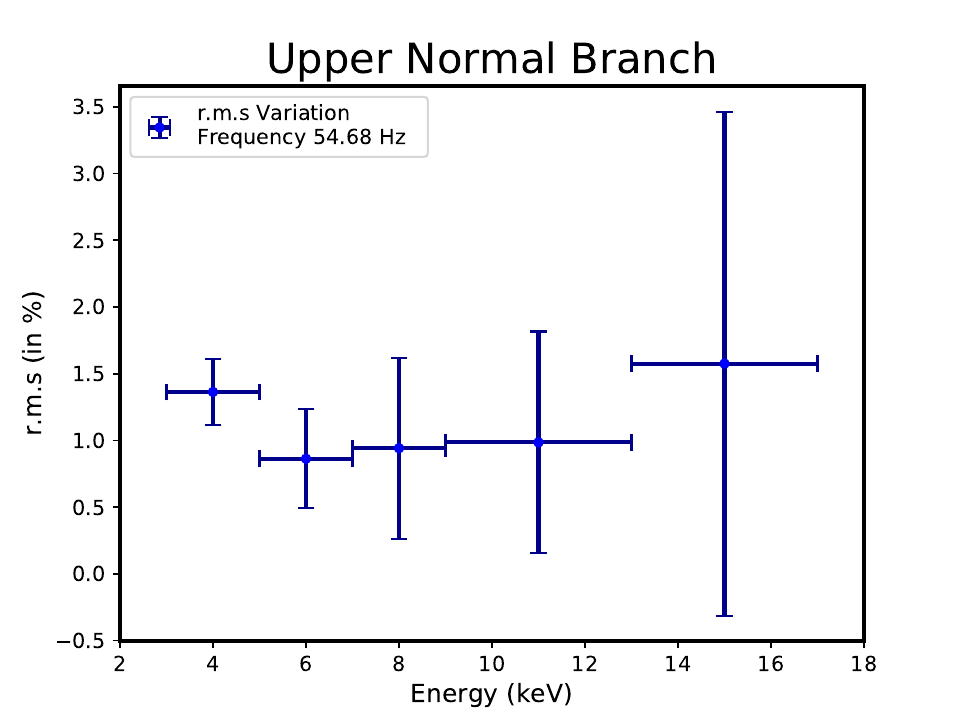}\\
		\includegraphics[width=0.32\linewidth]{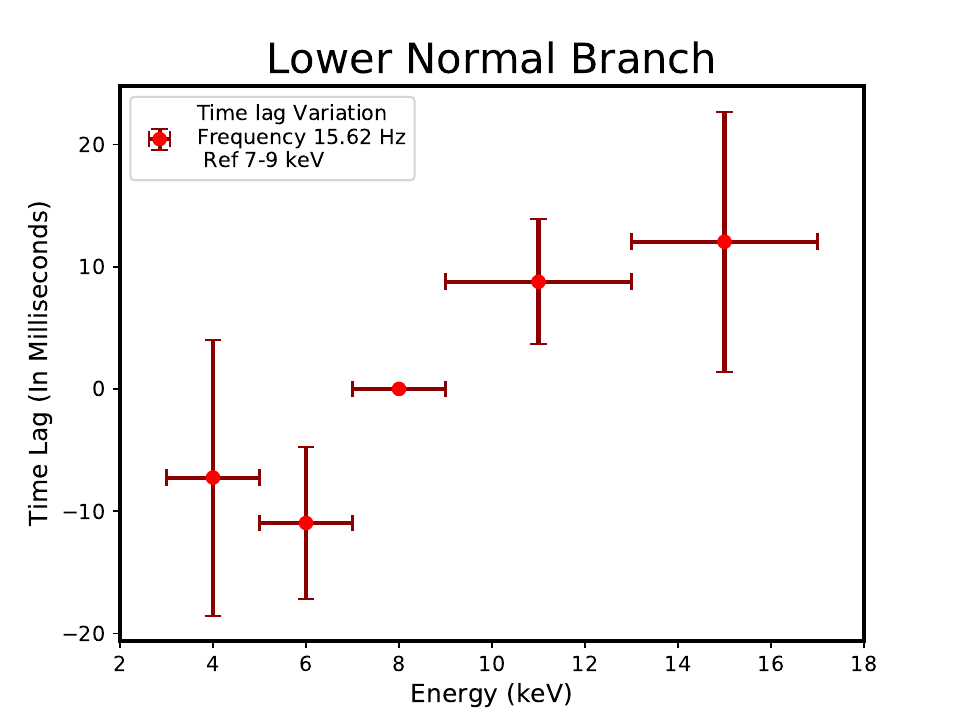}&
		\includegraphics[width=0.32\linewidth]{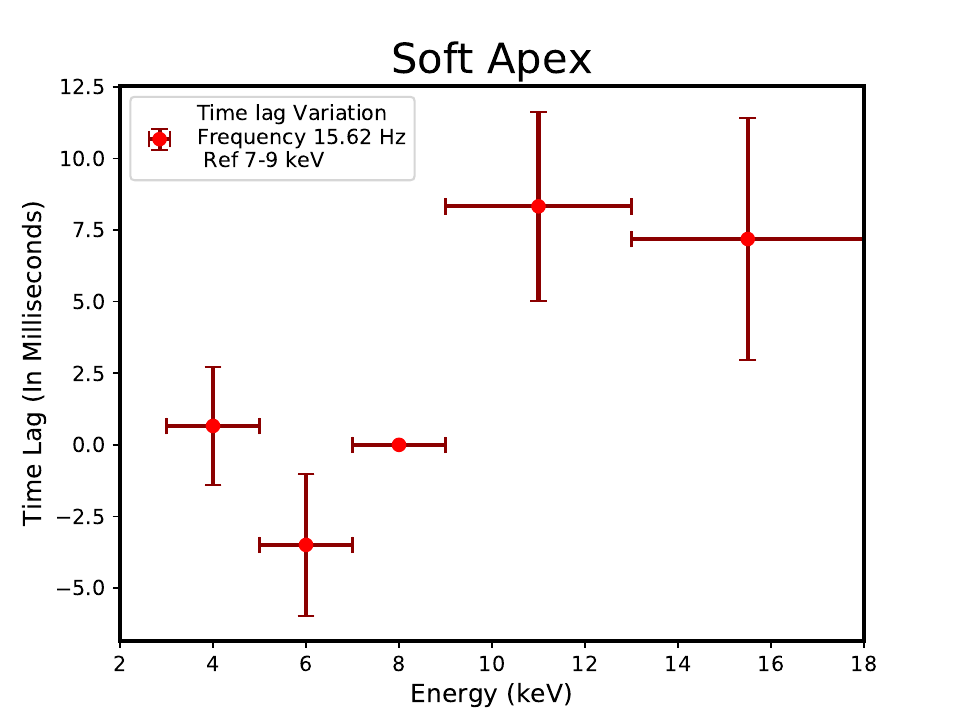}&
		\includegraphics[width=0.32\linewidth]{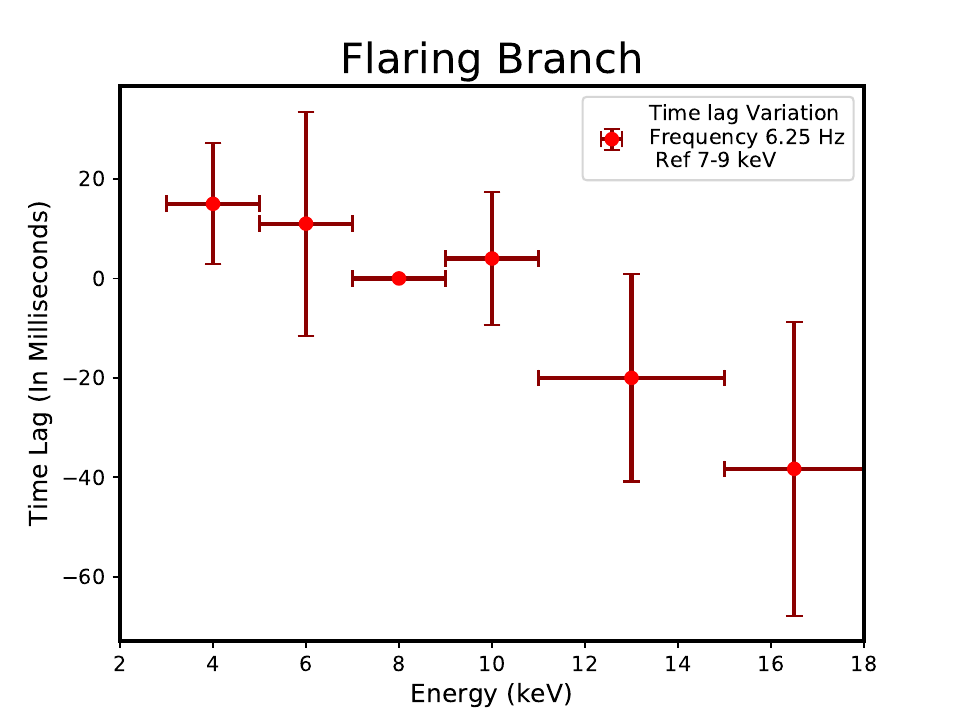}\\
		\includegraphics[width=0.32\linewidth]{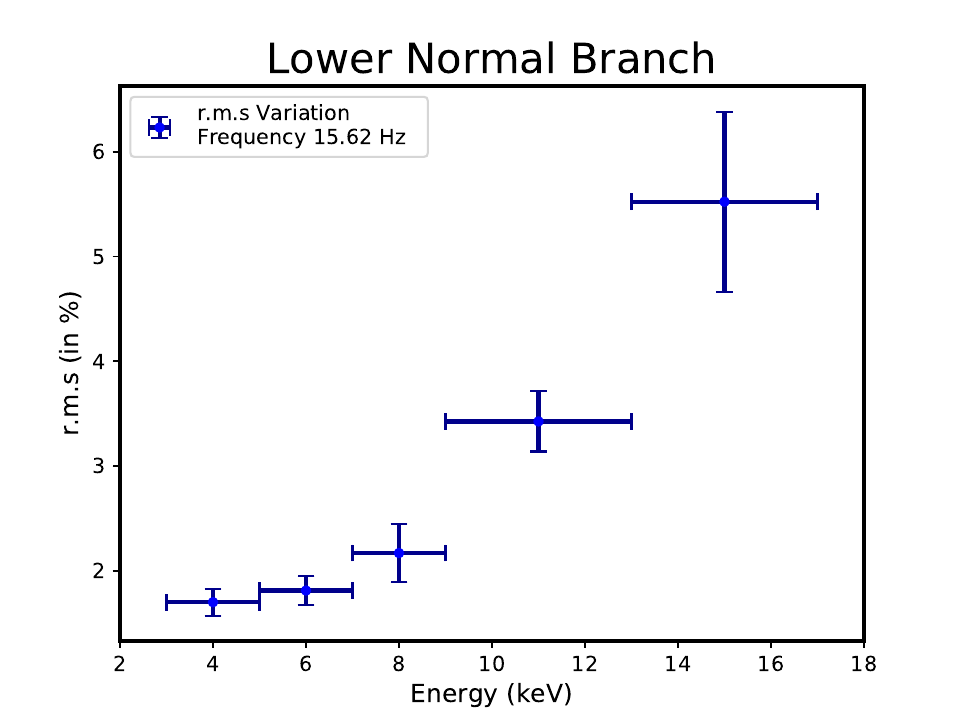}&
		\includegraphics[width=0.32\linewidth]{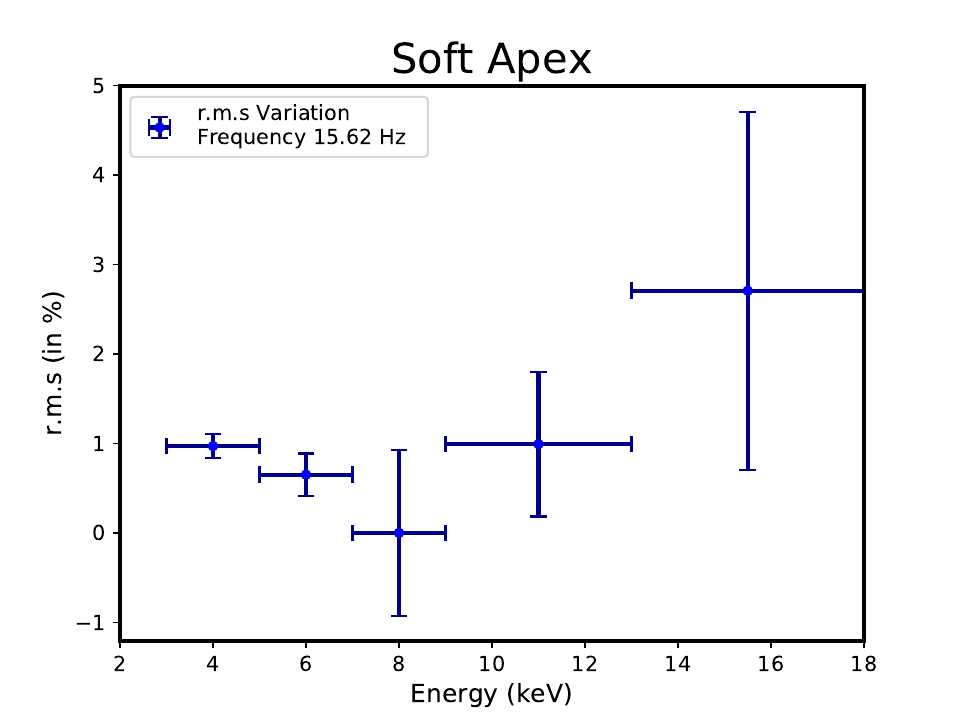}&
		\includegraphics[width=0.32\linewidth]{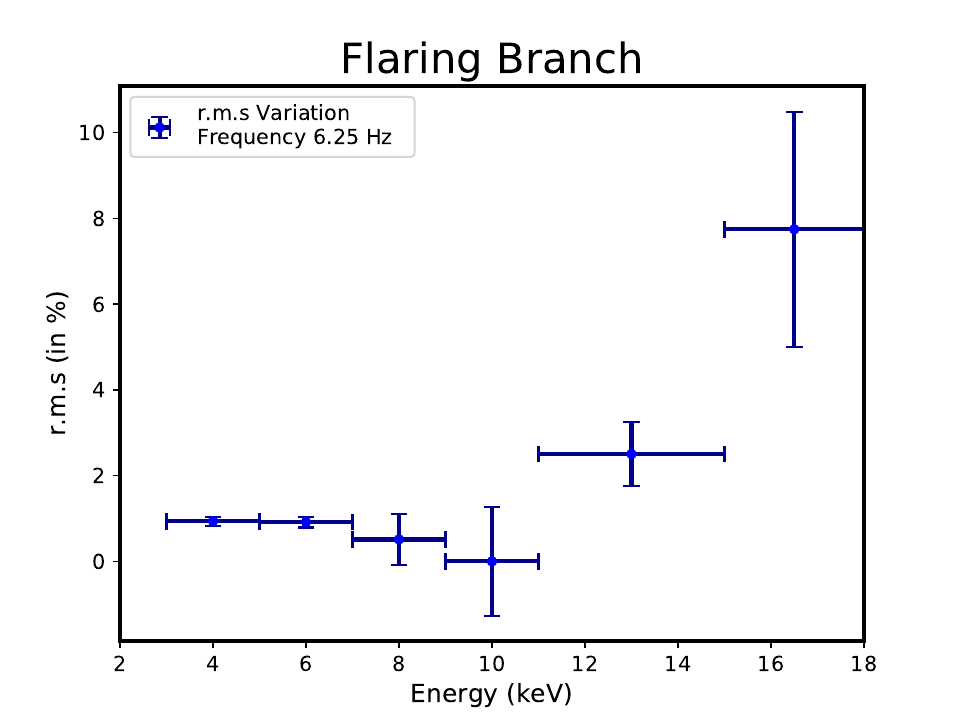}\\
		
	\end{tabular}
	
	\caption{Phase lag and fractional r.m.s amplitude variation with respect to energy in different segments using the reference energy band of 7-9 keV around the QPO frequency with in 3-20 keV. }
	\label{Fig8}
	
\end{figure*}

\section{Spectral Analysis}
Spectral analysis has been carried out for each of the six segments using strictly simultaneous data of both SXT and LAXPC 20 in the combined energy range of 1-25 keV (SXT 1-7keV, LAXPC 20 4-25keV).  For the spectral analysis,  LAXPC 20 was used since the detector has a lower background compared to the others. A gain correction of the fixed slope of unity has been applied to the SXT data. A constant factor was multiplied to the SXT data to take into account calibration uncertainties between the two instruments which turned out to be in the range of 0.85-0.98. A systematic error of $3$\% was added to take into account response uncertainties.\\
We initially fitted the spectra by considering that the photons are generated from an accretion disk around the neutron star entering a comptonized medium. Hence we have modelled the radiation by multi-coloured blackbody radiation from the disc using the {\bf Xspec} model {\tt diskbb} \citep{klitzing:1984PASJ...36..741M} convolved with the Comptonized model {\tt thcomp} \citep {klitzing:2020MNRAS.492.5234Z}. The absorption of the emitted radiation has been taken into account by the {\bf Xspec} routine {\tt tbabs}. Thus the {\bf Xspec} model format used was  {\tt const*tbabs*thcomp*diskbb}. Since  {\tt thcomp }is a convolution model, the energy range for model computation was taken from 0.1 to 200 keV with 500 logarithmic bins to fit the spectra in the 1 -25 keV range. For most of the segments, the fit was not acceptable with a reduced $\chi_{red}^2 \sim 2$ as shown in Figure \ref{Fig3}. Replacing the Comptonized component with a power-law one led to an even larger $\chi^2$. The inner disc radius can be estimated from the normalization of the disc component by  $\left(N_{dbb} = {{R_{km}^2}\over {D_{10}^2}} \cos i\right)$ \citep{klitzing:2019RAA....19..114B}, where the distance $D_{10}$ is measured in units of 10 kpcs and $i$ is the inclination angle i. Assuming the distance to the source to be  $11$ kpc \citep{klitzing:penninx1993radio} and the inclination angle as 35\textdegree \citep{klitzing:10.1046/j.1365-8711.2000.03443.x,klitzing:DAi:2008tiy}, the inner disc radius for different segments varied from 2-6 km which is rather unphysically small.\\
Next, we replaced the disc emission component with a blackbody one represented by the {\bf Xspec} model {\tt bbodyrad}. This yielded a much better fit with reduced $\chi^{2}_{red} \sim 1$ as shown in Figure \ref{Fig3}. Moreover, the inferred radius of the blackbody emission from the normalization $\left(N_{bb}= {{{R^2}_{km}}\over{D_{10kpc}^2}}\right)$ gave more physically acceptable values of $\sim 25$ km.\\ 
For all branches, error computation for the electron temperature and the optical depth of the thcomp model, showed that both of them were not constrained. Considering instead the electron temperature and spectral index as parameters, gave the same result. Fixing the electron temperature at 3 keV, the optical depth was constrained to be $\sim 10$ for the first four of the branches (HB, HA, UNB, LNB), however for the other two branches (LNB and FB) it was still unconstrained. If instead, the optical depth was fixed at $10$, then for all branches the temperature was well constrained and it is this set of fitting which is reported in Table \ref{Table2}. As an example, the unfolded spectra for the soft apex and residuals are shown in Figure \ref{Fig4}, while Table \ref{Table2} lists the best-fit parameter values with errors computed at 90\% confidence levels.\\
Table \ref{Table2} also lists the unabsorbed total flux using the {\bf {\bf Xspec}} function {\tt cflux} for the total and blackbody flux with in energy band of 0.01-100 keV. The Compton flux is defined as the subtraction of the blackbody flux from the total. The blackbody flux ratio is defined as the blackbody flux to the total, while the Compton flux ratio is defined as the Compton flux by the total flux.\\
The top right panel of Figure \ref{Fig5} shows the variation of fluxes with the blackbody flux ratio with the Z-track points labelled. It is clear from the plot that the blackbody flux ratio increases monotonically as the source moves along the Z-track i.e. from the horizontal branch to the flaring one. While the blackbody flux increases and then decreases, the Compton flux decrease along the track. Indeed, the variation of the Compton flux with the total flux gives a Z-track as shown in the top left panel of Figure \ref{Fig5}. The bottom left panel of Figure \ref{Fig5} shows the variation of the blackbody temperature and radius as a function of the blackbody flux ratio, while the bottom right panel shows the variation of the Corona temperature and covering fraction. Note that the covering fraction monotonically decreases along the track, giving rise to the decrease in the Compton flux.\\
The effect of the decrease in the Comptonized component flux as the source moves along the Z-track is further illustrated in Figure \ref{Fig6}, where the ratio of the model Comptonization count rate to the total counts (which is denoted as Compton Model Count Ratio) is plotted against energy. For low soft energies, the ratio significantly decreases as the source moves from HB to FB, while the hard X-rays remain dominated by the Comptonized component except for the FB spectrum.
\section{Timing Analysis}
The power density spectra (PDS) were computed using all three LAXPC in the energy range 3-20 keV using the LAXPC software ``{\tt laxpc\_find\_freqlag}'' and rebinned using "{\tt laxpc\_rebin\_power}"\footnote{\url{http://astrosat-ssc.iucaa.in/uploads/threadsPageNew\_LAXPC.html}}. The PDS were generated using lightcurves with a bin size of 0.25 millisecond corresponding to a Nyquist frequency of 2000 Hz. The lightcurve was split into segments of time length  8.192 sec and then the PDS of the segments were averaged. Thus the minimum frequency of the PDS is 0.121 Hz. The PDS were then further rebinned in frequency.
The software subtracts out the Poisson noise contribution assuming a dead time of 42 microsecond. However, for some of the PDS, we find a faint residual constant contribution, which we model as a  constant.\\
Figure \ref{Fig7} shows the PDS for all six segments which have been fitted with multiple Lorentzian to take into account the broad band continuum and QPO features. The HB, hard apex, and upper normal branch show a prominent QPO at $\sim 40$ Hz, while the lower normal branch, soft apex, and the flaring one show a feature at $\sim 7$ Hz. There is a harmonic of the QPO in the soft apex PDS. The best fit Lorentzian parameters are tabulated in Table \ref{Table3}.\\
To quantify the energy dependence of the QPOs, we estimated the fractional r.m.s. and the phase-lag at different energy bands using the LAXPC software  {\tt laxpc\_find\_freqlag}. The software estimates the fractional r.m.s and the phase-lag for a frequency bin centered at a particular frequency and a bin width $\Delta \nu$, which have been chosen to roughly correspond to the frequency and the width of the QPO. The frequency used for the computation and the bin width $\Delta \nu$ have been listed in Table \ref{Table4}.\\ 
Figures \ref{Fig8} show the fractional r.m.s (rms amplitude) and the phase lags computed for different energy bands for all the six QPOs observed in the different sections. For the phase lags, the reference energy band is 7 - 9 keV. 
For all the QPOs the rms amplitude increases with energy suggesting that the Comptoinized emission plays a dominant part in the phenomenon. For four of the segments (HB, HA, UNB, and FB) the system seems to exhibit soft phase lags i.e. the soft photons lag the harder ones, while for the other two (SA and LNB) it seems to be the opposite.\\
A model for such QPOs is that there is an inner hot flow surrounded by a truncated accretion disc, and the inner hot flow precesses due to Lense-Thirring precession \citep{10.1111/j.1745-3933.2009.00693.x, 2016MNRAS.461.1967I}. Due to the precession, the inclination angle to the observer of the inner flow changes and hence there is a variation of the  thermal Comptonized spectrum which gives rise to the observed QPO. In this framework, the energy dependent r.m.s and time lag arise due to spectral changes as the inner region precesses, but quantitative predictions are difficult .\\
Support for the model comes from the observed dependence of the nature of the QPO time lags with the inclination angle of the source \citep{klitzing:10.1093/mnras/stw2634, klitzing:2019A&A...625A..90R} and the variation of the broad Iron line energy with QPO phase \citep{klitzing:10.1093/mnras/stu2048}. However, more direct evidence such as the expected dependence of the QPO frequency with the truncated disc radii, has not been forthcoming \citep{klitzing:10.1093/mnras/stab3803}. The discovery of simultaneous $\sim 50$ Hz and kHz QPOs in a slowly rotating pulsar in Terzan 5 \citep{klitzing:Altamirano_2012} implies that the Lense-Thirring  model is not universally applicable to all systems.\\
The QPO may also be due to variations in the heating rate of the corona. The time lag between different energy bands can then be identified with the light crossing time of the photons as they scatter in the Comptonizing medium. While the process generically produces hard lags, soft lags can occur if a fraction of the Comptonized photons impinge back to the seed photon source and the model explains the energy dependent r.m.s and time lags observed for khz QPO \citep{klitzing:LeeH.C.MisraR.TaamR.E.2001, klitzing:kumar2014energy,klitzing:kumar2016constraining,klitzing:karpouzas2020comptonizing}. The same model has also been successfully applied to explain the r.m.s and time lags of lower frequency QPOs \citep{klitzing:10.1093/mnras/stab827,klitzing:10.1093/mnras/stac1202,klitzing:10.1093/mnras/stac1922}. The spectral fitting reported in this work, shows that the covering fraction of the Comptonizing cloud is significantly less than unity and that in the energy range $> 4$ keV, there is significant contribution by the seed blackbody component. On the other hand, the presently available codes \footnote{\url{https://github.com/candebellavita/vkompth}} assume that the emission is entirely due to Comptonization and hence we cannot use them to  directly fit the energy dependent r.m.s and time lags. Nevertheless, qualitatively, one can infer that the soft time lags observed in most of the branches indicate that a significant fraction of the Comptonized photons impinge back to the seed photon source, while for the LNB and SA, the hard lags suggest that this fraction is reduced. Scaling from earlier reports, that $\sim 50$ microseconds lags are obtained from a corona size of $\sim 10$ km \citep{klitzing:kumar2016constraining,klitzing:karpouzas2020comptonizing} whereas the size of the corona should be $\sim 150$ km for the $\sim 50$Hz QPOs and $\sim 10^3$ km for the few Hz ones to produce the observed time lags. Since most of the gravitational energy is released close to the neutron star, an efficient energy transfer mechanism is required to heat the corona to such distances. A complex geometry of such a corona is required since the covering fraction is inferred to be $\sim 0.5$. Earlier reports of simultaneous detection of kHz and 50 Hz QPOs \citep{klitzing:Ford_1998,klitzing:Wijnands_1997} in such systems, would again point to a complex geometry with the two QPO having different scale lengths.\\
Stochastic variations produced locally in different radii of the disc, can propagate inwards leading to the broadband variability observed in X-ray binaries \citep{klitzing:10.1093/mnras/292.3.679, klitzing:10.1093/mnras/stz930}. Since the inner region (either due to higher temperature of the inner disc or a hot inner corona) produces harder X-rays than the outer, such propagations can lead to time delays on the viscous time-scales \citep{klitzing:2001MNRAS.327..799K} although further investigation is needed as  \citet{10.1093/mnras/stx991} identified seignificant discrepancies between the observed lag in the BHLMXB source XTE J1550-564 and model predictions. The time delays could also be related to the sound crossing time and would be frequency dependent \citep{klitzing:2000ApJ...529L..95M}. A particular scenario is when fluctuations in the inner edge of a truncated disc propagate to the inner hot flow after a time delay, which has been used to explain the r.m.s and time lags of the broadband noise in the black hole system Cygnus X-1 \citep{klitzing:10.1093/mnras/stz930}. A similar interpretation has also been undertaken for QPOs \citep{klitzing:10.1093/mnras/staa2506, klitzing:10.1093/mnras/stac1490}, where the QPO is caused by a variation of a physical parameter of the truncated disc (such as radius or temperature) which leads to a variation in a physical parameter of the inner flow (such as heating rate, optical depth, covering fraction) after a time delay. Alternatively, the QPO could be caused by a variation in the corona and the disc could respond after a time delay. Using numerical differentiation of spectra with respect to physical parameters, the model can predict energy dependent r.m.s and time lags which can in principle be fitted to good quality observations, to constrain which pair of parameters are varying and the time lag between them. While not presenting a complete picture of the phenomenon, the analysis does provide useful insights, albeit within the framework of the model.  However, for the quality of the data reported in this work (especially the large errors on time lags), a full statistical analysis is not warrented since the parameters will not be well constrained and several pairs of them will provide similar fits. The expected rms amplitude and phase-lags will then be a complex function of several parameter variations and would need to be solved numerically. Given the quality of the data, this may not be warranted since the data would not be able to constrain a large number of free parameters of the complex model. Nevertheless, such endeavours, using wider band data (e.g. from {\it NICER} and {\it AstroSat} ) may unlock the radiative nature of the QPO phenomena.\\ 

\section{Summary \& Discussion}
Using  {\it{AstroSat}}'s  SXT and LAXPC data, we have studied the complete spectral evolution of the source in the 1-25 keV band as GX 340+0 moved along its Z-track which we divided into six distinct sections, the Horizontal brand (HB), the Hard Apex (HA), Upper Normal branch (UNB), Lower Normal branch (LNB) and the Flaring branch (FB).\\
The spectra of all segments are well represented by a model where a blackbody source is partially covered by a Comptonizing hot corona, while a model consisting of a corona Comptonizing disc emission, results in an unacceptable fit and unphysical inner disc radii of $\sim 2$ km. The Compton flux (defined as the total flux minus the blackbody one) decreases monotonically as the source transitions along the Z-track from HB to FB. So does the covering fraction of the corona which decreases from $\sim 0.4$ to $\sim 0.04$.   The ratio of the blackbody flux to the total flux (blackbody flux ratio) decreases along the track from $\sim 0.8$ at HB to $\sim 0.98$ at FB. The temperature of the corona also seems to decrease along the track, but there seems to be an increase for the FB, however, the coronal temperature estimation depends on our assumption that the optical depth of the corona is a constant. All of these indicate that as the source evolves the Comptonization component decreases monotonically.\\
The Z-track shape is caused by the complex evolution of the blackbody component whose flux increases from HB to HA/UNB and then decreases for LNB/SA and finally increases at the FB.  It is this non-monotonic behaviour of the blackbody component which causes the source to trace out the Z shape curve in its HID. In fact, the Z-shape behaviour can be seen when one plots the Compton flux versus the total flux (Top right panel of Figure \ref{Fig5}).The temperature of the blackbody also follows more or less the same complex behaviour.The variation of the black body flux is primarily due to variations of the temperature. Moreover, The black body radius seems to be somewhat inversely correlated with the temperature.The increase of the black body temperature along with a  reduction of the radius may describe the condition for the formation of the jets \citep{klitzing:2006A&A...460..233C}, since that may imply that the corona is becoming more radiation pressure dominated. Note that the contraction of the radius in the flaring branch has been previously reported for the Sco-like sources \citep{klitzing:Church:2012nb} and for X 1624-490  \citep{klitzing:2001A&A...378..847B}  but not for the Cyg-like sources.\\ 
The nature of the high-energy spectra shows a dramatic evolution as the source evolves. In the 10-20 keV band, the blackbody component is only about a few percent of the flux in the HB, but as the source evolves the blackbody component is about 90\% of the flux in the FB. Thus in this energy band, the source moves from being mostly Compton emission during HB  to being blackbody dominated during FB.\\
Rapid timing analysis reveals that the power density spectra in HB, HA, and UNB are dominated by a $\sim$ 50 Hz QPO whose centroid frequency increases from $\approx$ 41 Hz to 50 Hz. On the other hand for LNB, SA, and FB there is prominent QPO at $\sim 6$ Hz. The fractional rms amplitude of these QPOs increase with energy. The phase lags seem to show complex behaviour with the lag being soft (i.e. the low-energy photons are delayed compared to the high-energy ones) for HB, HA, UNB, and FB but are reversed for the LNB and SA branches.\\
Our results show the potential of undertaking broad-band spectral and temporal analysis of neutron star systems and highlight the need for coordinated observations between {\it NICER} and {\it AstroSat} which would provide spectral and rapid timing information in a wide energy band. Analysis of such data with more sophisticated spectro-temporal models would reveal the nature of these sources.
\section{acknowledments}
In this research work, we have used the data from LAXPC and SXT payloads onboard {\it  AstroSat} available at ISSDC(Indian Space Science Data Centre). We are thankful to ASC (AstroSat Science support cell ) for helping us by providing their support. SC is thankful to IUCAA (Inter University Centre for Astronomy and Astrophysics) for providing the periodic visit to carry out the major part of the research. We are thankful to the LAXPC Payload Operation Center (POC) and SXT POC at TIFR, Mumbai. We would also like to thank the members of the LAXPC and SXT instrument teams for their contribution to developing the instrument. We have made use of the software provided by the HEASARC (High Energy Astrophysics Science Archive Research Center ). Furthermore, SM would like to thank IUCAA for the visiting associateship. The research leading to these results has been funded by the Department of Space, Govt. of India, ISRO under grant no. DS\_2B-13013(2)/10/2020-Sec.2.SM also acknowledges SERB for the research grant CRG/2019/001112.
\section{Data Availability}
The LAXPC and SXT archival data that has been used in this article can be found at {\it AstroSat} ISSDC website ({\url{https://astrobrowse.issdc.gov.in/astro\_archive/archive}).
\bibliographystyle{mnras}
\bibliography{paper} 
\appendix

\bsp	
\label{lastpage}
\end{document}